\documentclass[aps, twocolumn,showpacs,amsmath,amssymb,pre, longbibliography]{revtex4-1}

\usepackage{microtype}
\usepackage{graphicx}
\usepackage{subfigure}
\usepackage{booktabs} 
\usepackage{todonotes}
\usepackage{graphics,graphicx}
\usepackage{xspace}
\usepackage{dsfont}
\usepackage{bm}
\usepackage{mathtools}
\usepackage{amssymb}
\usepackage{mathtools}
\usepackage{tcolorbox}
\usepackage{float}
\usepackage{amsthm}
\usepackage{hyperref}

\newcommand{\ui}{\textrm{i}}
\newcommand{\figref}[1]{~Fig.\ref{#1}}

\newcommand{\eqreff}[1]{~Eq.\eqref{#1}}

\newtheorem{definition}{Definition}

\hypersetup{
	colorlinks,%
	citecolor=blue,%
	filecolor=blue,%
	linkcolor=blue,%
	urlcolor=blue
}
 
\begin{document}
	\title{Beyond symmetrization: effective adjacency matrices and renormalization for (un)singed directed graphs }




	\author{Bruno Messias}
	\email{devmessias@gmail.com}
	\affiliation{São Carlos Institute of Physics
		University of São Paulo,
		São Carlos, SP, Brazil}
	\thanks{This work was carried out during the \href{https://www.teses.usp.br/teses/disponiveis/76/76132/tde-06112023-105626/pt-br.php}{author’s PhD} program at the University of São Paulo (USP), São Carlos, Brazil. }

\begin{abstract}
	To address the peculiarities of directed and/or signed graphs, new Laplacian operators have emerged. For instance, in the case of directionality, we encounter the magnetic operator, dilation (which is underexplored), operators based on random walks, and so forth. The definition of these new operators leads to the need for new studies and concepts, and consequently, the development of new computational tools. But is this really necessary? In this work, we define the concept of effective adjacency matrices that arise from the definition of deformed Laplacian operators such as magnetic, dilation, and signal. These effective matrices allow mapping generic graphs to a family of unsigned, undirected graphs, enabling the application of the well-explored toolkit of measures, machine learning methods, and renormalization groups of undirected graphs. To explore the interplay between deformed operators and effective matrices, we show how the Hodge-Helmholtz decomposition can assist us in navigating this complexity.
\end{abstract}
		\maketitle

	In social networks the relationships between members commonly are asymmetric. This asymmetry means that if some user of a given platform follows another user the contrary is not necessarily true. These relationships can also have a negative value representing for example, a distrust of a citizen about a politician or a dislike towards a given topic
	\cite{zhengSocialNetworksRich2017a}.
In the field of sociometry, the asymmetry with positive or negative relationship between members of social groups can be represented by a data structure known as directed signed graph
	\cite{Harary_1953}.
	In biology, the same kind of data structure is used to represent protein-protein interactions\cite{aittokallioGraphbasedMethodsAnalysing2006};
where the negative and positive directed edges represent the activation and inhabitation between the biological elements. Therefore, the development of techniques aiming the characterization of signed digraphs  has great relevance today, for example, to identify criminal groups \cite{zhengAnalysisCriminalSocial2015},
	and anomaly behaviors in social networks
	\cite{Li_Yuan_Wu_Lu_2018}
and to help uncover protein-protein interactions. 

In the context of non-signed undirected graphs, a myriad of spectral methods and machine learning techniques related  to graph operators have been developed in order to analyze real world networks. The vast majority of these methods reside in two properties about the combinatorial adjacency and Laplacian operators: the existence of an orthonormal basis for the operators and the fact that the eigenvalues associated with them residein the real line. 
Unfortunately, in the case of signed digraphs these two properties are usually absent \cite{Li_Yuan_Wu_Lu_2018}.

Due to the absence of these two properties of the combinatorial operators the approach of deformed Laplacian operators\cite{fanuelDeformedLaplaciansSpectral2019} is a promising approach. In this approach, we have group deformations that can be used in the combinatorial Laplacian to study specific properties. For example, in the context of directed edges we can talk about hierarchy and circularity. The hierarchy in real networks is related to the ranking problem, which aims to associate to each vertex a ranking score in order to find hidden hierarchical groups in the associated digraph
	\cite{shahriariRankingNodesSigned2014,gemiciVisualizingHierarchicalSocial2018}.Recently\cite{debaccoPhysicalModelEfficient2018}
	the authors proposed a new raking technique called SpringRank obtained from a variational principle. The SpringRank has strong relationship  to
	the HodgeRank. The Hodgerank is a powerful technique proposed by Jiang et. al. in 
	\cite{jiangStatisticalRankingCombinatorial2011} based
	in the Hodge-Helmotz decomposition theorem. It finds various applications in sociometry\cite{hodge2019br, hodge2024br}, network dynamics\cite{hodgeGowithTheFlow,hodgeDynamics}, and analysis of financial transactions\cite{hodgebitcoin}, to name just a few.  Remarkably, the ranking scores obtained from the first eigenvector of the deformed by dilation Laplacian constitutes a good approximation for the HodgeRank when the dilation parameter is small enough\cite{fanuelDeformedLaplaciansSpectral2019}.

	The circularity property is related to the presence of cycles; a triad is an example of a cycle which has been extensively used in order to define the notion of balance in social systems
	\cite{hararyNotionBalanceSigned1953,arefMultilevelStructuralEvaluation2020}
	and  to improve
	the algorithms aiming to analyze
	social networks datasets\cite{Chen_Qian_Liu_Sun_2018}.
In this work, we interpret circularity as the degree of inconsistency introduced by directed edges. To measure this inconsistency, we employ a formalism originally developed for studying the movement of a quantum charged particle in a discrete space affected by magnetic perturbation  \cite{shubinDiscreteMagneticLaplacian1994}.
 Historically motivated, this approach is known as the magnetic Laplacian.
 
  Recently, the magnetic Laplacian has found applications in unsigned digraphs for refining community detection algorithms \cite{fanuelMagneticEigenmapsCommunity2017},
	characterizing and comparing networks\cite{chaos2020,complexNetWeights}, examining spectral properties of random matrices\cite{peronSpacingRatioCharacterization2020},improving graph machine learning algorithms\cite{magnetNeural}, 
	and analyzing tabular datasets\cite{tab2graph}. Surprisinly, this magnetic operator
	can been seen as deformation of the Laplacian by a $U(1)$ frustration\cite{fanuelDeformedLaplaciansSpectral2019}.

	These advancements in deformed Laplacian operators, while innovative, necessitate new computational and analytical frameworks, prompting critical inquiries: When should one deformation be preferred over another? Can the deformation framework be integrated with conventional tools from the analysis of undirected, unsigned graphs?

 Addressing the first inquiry, the application of the Hodge-Helmholtz decomposition was used as a strategy for directed graphs. This technique decomposes a directed graph into gradient, curl-free, and harmonic components, offering a way to analyze how some quantities defined on deformed Laplacians approaches behave in each decomposed graph. Such an analysis not only underscores the necessity for cautious application of deformation operators but also advocates for the utilization of effective matrices as a means to simplify and clarify the complex dynamics inherent in directed graphs.

In response, this work introduces the concept of effective adjacency matrices. This approach simplifies the analysis of directed and/or signed graphs by transforming them into their unsigned, undirected counterparts beyond just symmetrization of the adjacency matrix. This transformation leverages the established toolkit for undirected graphs, including centrality measures, clustering algorithms, and machine learning techniques.

	\section{Methodology}
\subsection{Group deformations in the combinatorial Laplacian}

Consider a graph $G=(V, E, w)$, an undirected, unweighted and unsigned graph defined by the set of vertices $V$, the set of edges $E$ formed by unordered pairs $(u, v)$ where $u, v \in V$; the edges of a graph can be represented by the adjacency $A$, and , $w: E \mapsto \mathbb{R}_{+}$ is the weight function that can be represented by the matrix $\mathbf{W}$.

We denote by $L_V$ the space of functions $f: V \to \mathbb{R}$ equipped with the following inner product:

\begin{align}
\langle f, g \rangle_V = \sum\limits_{v\in V} f(v)g(v)
\end{align}

The combinatorial Laplacian operator is a map $L: L_V \to L_V$, which is well-known and is given by:
\begin{align}
(L f)(u) = D(u)f(u) - \sum_{v\in Nei(u)} w(u, v) f(v),
\label{eqLaplacianCombinatorial}
\end{align}
where $\text{Neigh}(u)$ is the neighborhood of vertex $u$ and $D(u) = \sum\limits_{v \in \text{Neigh}(u)} w(u, v)$ is its degree.

Let $\mathcal{V}$ be a normed vector space with norm $\parallel \dot \parallel_\mathcal V$  and identity $\varepsilon$.
Let $\mathcal{G}$ be a group that acts on $\mathcal{V}$. The fundamental ingredient of group deformations is the existence of a flow on the edges of the type $\psi: E \to \mathcal{G}$ with the additional constraint that $\psi_{uv} = \psi^{-1}_{vu}$.
 Clearly, respecting the constraints, the choice of $\mathcal{G}$ is arbitrary. However, in the case of directed graphs and/or  signed graphs, a $\psi$ and consequently a $\mathcal{G}$ emerge related to the flow defined by the decomposition of the weight function for example $A-A^T$. We will return to this point later. In the current work, we will restrict ourselves to the case of groups whose representations in $\mathcal{F}$ are matrices in $C^{|V| \times |V|}$.

\subsection{Generalized degree of a vertex}
\begin{figure*}
	\centering
	\includegraphics[width=0.7\textwidth]{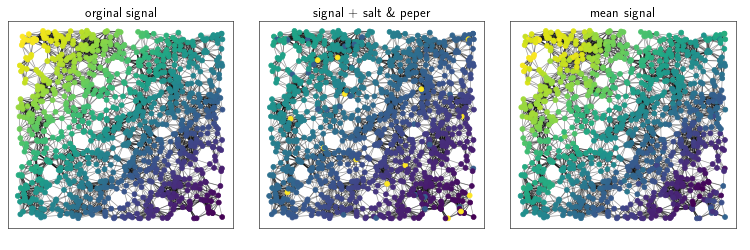}
	\caption{The left image shows a signal on a geometric graph. The center image displays the same graph with the signal contaminated by salt \& pepper noise. The right image shows the result of filtering using the aggregation of discrepancies between the signals, that is, $f(u) = f(u) + \frac{1}{D(u)}\sum_{v\in Nei(u)} (f(v) - f(u))$. }
	\label{figSignalDidatic}
\end{figure*}

Aggregating a signal with respect to its neighbors can be  used to reduce certain types of noise in images and graphs, as illustrated in  \figref{figSignalDidatic}. In a way, we can think of the degree of a vertex in a graph as a weighted sum of a constant function, $f$, over the neighbors of that vertex. This motivates us to define in this work a generalized definition for a degree of a vertex. To do so,  let be  an $T \in \mathcal{G}$ with a representation in $\mathcal{F}$. We define the promotion, $P$, as
\begin{align}
	P_{uv}^{tz} = T_{vu}^* T_{tz}.
\end{align}

\begin{definition}[Generalized Degree Function]
	Let $H: V \times V \to \mathcal{V}$, and let $f: V \to \mathcal{F}$ be a  function defined on the vertices.
	We define a generalized degree function $D: \mathcal{H} \times V \to \mathcal{V}$ such that
	\begin{align}
	 D_{\mathcal G}(H, f, u)
	= \nonumber\\
	\sum\limits_{v\in Nei(u)}
	w(u, v)(
	P_{vu}^{uv}H(u, v)f(v)- P_{uv}^{uv}H(u, u)f(u))\nonumber\\
	= \sum\limits_{v\in Nei(u)} w(u, v)(\omega_H(u, v)f(v) - w_{H,\mathcal G}(u, v)f(u))
	\label{eqGeneralizedDegree}
	\end{align}
	Note that if $\mathcal{V}$ is $\mathbb R$, $T=1$, and $h(u, v) = 1 - \delta_{uv}$, then the above function is nothing more than the well-known vertex degree.
\end{definition}

In matrix notation, let $\mathbf{F}$ be a vector representing a signal on the edges. We can associate the following matrix with it:
\begin{align}
\mathbf D_\mathcal G(H)  &=\mathbf W \odot( \mathbf{\tilde P} \odot \mathbf H - \mathbf D^{(P)}  ) \nonumber\\
&= \mathbf W \odot \mathbf  H_d,
\end{align}
where
\begin{align}
\mathbf D_{uv}^{(P)}=\begin{cases}
\sum\limits_{s \in Nei(u)}P_{us}^{us}h(u, u), \textrm{ se } u=v\\
0 \textrm { otherwise }
\end{cases}
\end{align}

A special case arises when $\mathbf{H} = \mathbf{A}$ and if $T_{uv} = 1$ we get back to the definition of usual degree, where $\mathbf{D}_{\mathcal{G}} = \mathbf{W} \odot \mathbf{\tilde{P}}$.

\begin{definition}[Deformed Laplacian]
	The definition of the generalized degree allows us to express the deformed Laplacian\cite{fanuelDeformedLaplaciansSpectral2019} as follows:
	\begin{align}
	(L_\mathcal G f)(u) &= D_\mathcal G(A, u)f(u) - \sum_v w_\mathcal G(u,v)f(v)
	\label{eqLaplacianDeformed}
	\end{align}
	$A$ is the adjacency matrix of $G$.
\end{definition}

The definition of a generalized degree given by~Eq.\eqref{eqGeneralizedDegree} was essential to express the deformed Laplacian ~Eq.\eqref{eqLaplacianDeformed} in a similar way to the combinatorial Laplacian ~Eq.\eqref{eqLaplacianCombinatorial}.

The generalized degree in relation to a vertex $u$ can be understood as the accounting of discrepancies of a signal ($f$)at a vertex
 $u$ in relation to its neighbors. This discrepancy is represented by the term $ \omega_{H}(u, v)f(v) - w_{H,\mathcal G}(u, v)f(u)$. We will now analyze the case where $\mathbf H = \mathbf A + \mathbf I_{|V|}$.In this specific case, we will omit the subscript $H$ from $\omega_H$ and $w_{H, \mathcal G}$.

\subsection{Effective graphs associated with group deformations}
\label{sec:methEffectiveMatrix}

For any given edge connecting two vertices
the discrepancy is calculated as the difference between the adjusted signals in the vertices. This difference is then assessed in terms of its magnitude (norm) in the vector space providing a quantitative measure of how ``discrepant'' or ``unalike'' the signals at the two vertices are.
\begin{definition}[Edge Discrepancy]
	We define the discrepancy on an edge as $\xi: E\mapsto \mathbb R_+$ for a given $f\in \mathcal V$
	as
	\begin{align}
	\xi_f (u, v) = \parallel
	\omega(u ,v) f(v) - w_\mathcal G(u, v)f(u)
	\parallel_\mathcal V
	\end{align}
\end{definition}

\begin{definition}[Generalized frustration]
	\begin{align}
	\eta_f = \frac{1}{
	2
	}
	\sum_{(u,v) \in E }
	\frac{w(u, v)\xi_f(u, v)^2}{
		\sum\limits_{u \in V}
		D_\mathcal G(A, u)\parallel f(u)\parallel_F^2
	}
	\end{align}
\end{definition}

As mentioned earlier, if
 $P_{uv}^{vu} = \varepsilon $ the degree function at $u$ with $H=A$
 is identical to the degree of vertex $u$ in the undirected graph. We will also have
$\xi_f(u, v) = f(v) - w_\mathcal Gf(u)$, if $\parallel f(u)\parallel^2_F = 1$ then the generalized frustration is equivalent to the frustration used in the group synchronization problem\cite{angSyncEigen,fanuelMagneticEigenmapsVisualization2018}  given by

\begin{align}
\frac{1}{2}
\sum_{(u,v) \in E }
\frac{w(u, v) \parallel f(v) - w_\mathcal Gf(u)\parallel^2_F  }{
	\sum\limits_{u \in V}
	\mathrm{Vol}(G)
}
\label{eqSync}
\end{align}
The generalized frustration provides a balanced measure of how much the graph edges deviate from being in harmony with each other when this graph has values, $f$, associated with the vertex.

We can use the definition of frustration to define a effective weight (not necessarily symmetric) as
\begin{definition}[Effective Weight]
	For a $\beta \in \mathbb{R}_+$
	\begin{align}
	w_e^{(f, \beta)}(u, v) = w(u, v)e^{-\beta\xi_f(u ,v)^2}
	\label{eq:effWeight}
	\end{align}
	Assuming for simplicity that  $\parallel f(u)\parallel^2_F = 1$ the generalized frustration can be written in terms of a logarithmic potential on the edges
	\begin{align}
	\eta_f = -\frac{\beta}{
		\mathrm{Vol}_\mathcal G(G)
	}
	\sum_{(u,v) \in E }
	w\ln \left(
			\frac{w_e^{(f)}}{w}
		\right)(u, v)
	\end{align}
\end{definition}

An obvious implication of the above definition is that for each undirected graph without signs equipped with a group potential, $G=(V, E, w, T)$ there is associated a family of graphs, $G_e^{(f, \beta)}=(V, E, w_e^{(f, \beta)})$, such that new characterizations and measures of
$G$ can be extracted by considering  $G_e^{(f, \beta)}$. In the case that
$P_{uv}^{vu} =1$ such graphs are undirected.

By reinterpreting edge weights through the lens of edge discrepancy and generalized frustration, we can effectively ``translate'' the  graphs into a format that adheres to the undirected model, thereby making them amenable to the vast toolbox of analytical techniques available for undirected graphs used in network science and machine learning on  graphs,

\section{Group deformations for directed graphs}

In the last years a very specific type of group deformation known as
magnetic Laplacian has been proposed and used to understand directed graphs
in a myriad of tasks such as community detection\cite{fanuelMagneticEigenmapsCommunity2017}, graph comparison and
characterization\cite{chaos2020} and machine learning techniques\cite{magnetNeural}. In essence
given a directed graph defined by a weight matrix $\mathbf W$  the magnetic Laplacian
commonly used  in literature starts decomposing this matrix into a symmetric and skew-symtric
part

$
    \mathbf W =
    \frac{\mathbf W + \mathbf W^T}{2} + \frac{\mathbf W - \mathbf W^T}{2}.
$
Then the skew-symmetric matrix, which we will associated with a
flux matrix $\mathbf A = \mathbf W - \mathbf W^T$,
is used to construct an edge-group potential given by
\begin{align}
    \mathbf T(q) = \exp(2\pi q \ui \mathbf A^T)
    \label{eqTmagnetic}
\end{align}
where $q \in [0, 1) $ is the charge parameter. The deformation given by the above equation implies that
$P_{uv}^{vu} =1$ and $\xi_f(u, v) = \xi_f(v, u)$.  It is self-evident that in this case
we are working with the $U(1)$ group.

Other type of group deformation  known as dilation Laplacian\cite{fanuelDeformedLaplaciansSpectral2019} can emerge when we choose the
following $T$-deformation
\begin{align}
    \mathbf T(\alpha) = \exp(\alpha \mathbf A^T)
    \label{eqTdil}
\end{align}
where $\alpha \in \mathbb R_{>0}$ is the dilation parameter. The dilation deformation
has a deep conection with the Hodge-rank technique as was remarked by\cite{jiangStatisticalRankingCombinatorial2011}.

An important point still unexplored in previous works is that there is no restriction to use
a flow matrix, $A$, coming from the skew-symmetric matrix $W - W^T$.
Both~Eq.\eqref{eqTmagnetic} and~\eqref{eqTdil} only have the restriction that the $A$ should be
asymmetric, i.e. $\mathbf A^T = -\mathbf A$. We can think in serveral cases that should be interesting
to explore for example using the google matrix instead of the adjacency matrix, random walks with
teleportation and other techniques. We will not explore these different properties in the present work.
\newline



	\section{Results}
\subsection{Interplay between Hodge-Helmholtz decomposition and group deformations}
In recent studies, the formalism of magnetic Laplacians has been extensively utilized for the characterization of directed networks. However, a question often neglected is whether the magnetic Laplacian and associated quantities invariably encapsulate the entirety of the information innate to the network's direction. Indeed, it has been demonstrated recently that the magnetic approach can yield less optimal results in certain directed graphs than simply ignoring the direction\cite{squirrelPaper}. Consequently, it becomes critical not only to highlight these limitations, but also to invest in research, techniques, and approaches that aim to unearth insights about the root cause of this behavior. One potential approach we propose and evaluate in this paper involves the utilization of the decomposition associated with the Hodge-Rank method. This method shares an intrinsic connection with dilation deformation, as we discuss in the methodology section and as presented in \cite{fanuelDeformedLaplaciansSpectral2019} work.

\begin{figure*}
	\centering
	\includegraphics[width=0.7\textwidth]{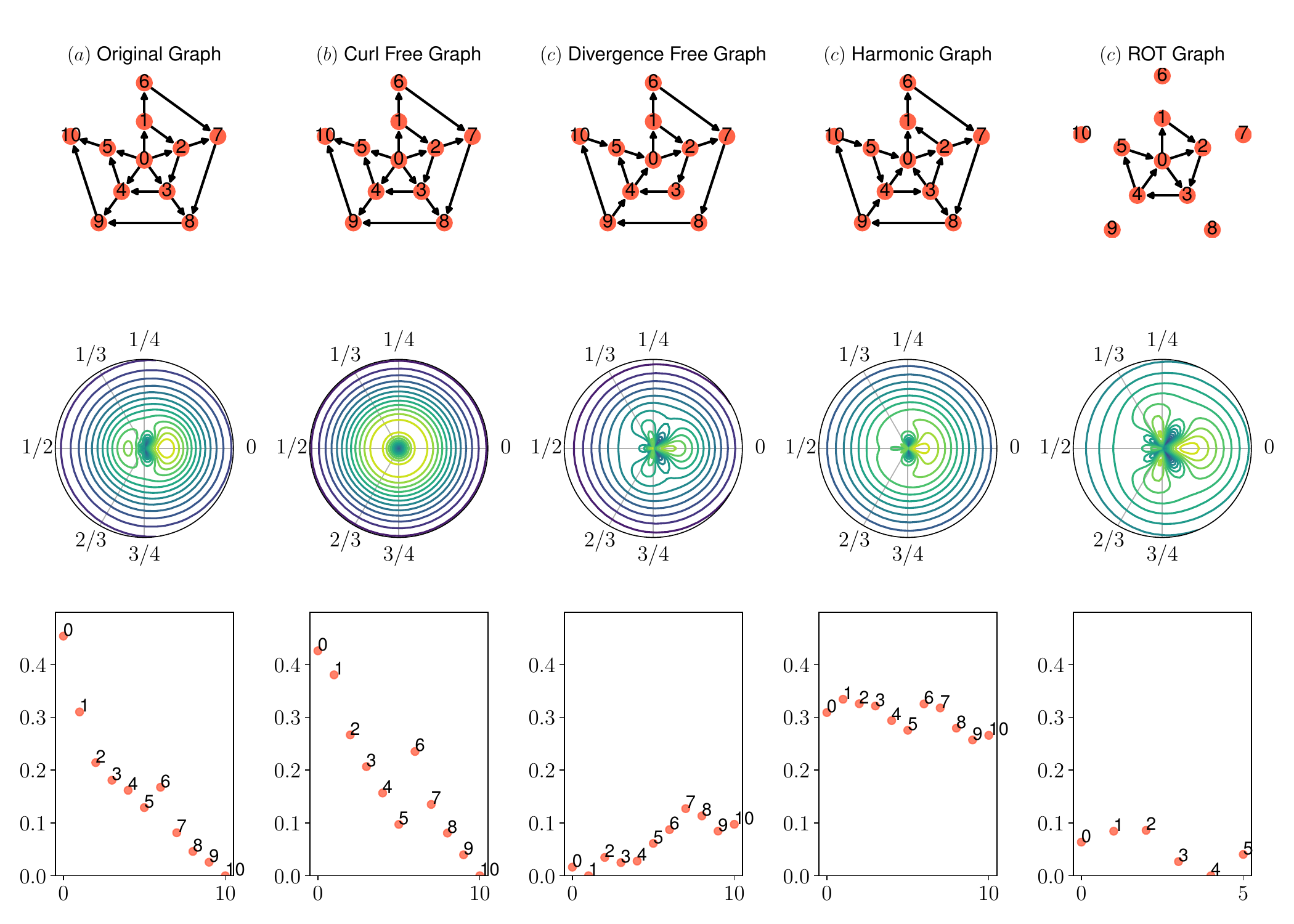}
	\caption{Helmholtz-Hodge decomposition of a directed graph and analysis of their respective magnetic and dilation Laplacians for each component. The top-left corner image presents a directed graph. The sequences illustrated in b, c, d, and e in the top row display the results of the HodgeRank decomposition of the  directed graph, represented in curl-free, divergence-free, harmonic, and rotational components respectively. The second row portrays the magnetic specific heat for the original graph and each of its components. Lastly, the score obtained by the dilation Laplacian is displayed in the final row, both for the original graph and for each of its components. Interestingly, in some components, the result of the magnetic Laplacian is informative while the dilation Laplacian is not, and vice versa.
	}
	\label{fig:hodgeDecompAndDeformations}
\end{figure*}

The decomposition facilitates the breaking down of the flow defined by the directed edges ($\mathbf A- \mathbf A^T$) into gradient, curl-free, and harmonic flows. Each edge within these flows associates with a real, positive, or negative value. If the edge has a negative value then the direction is flipped. We demonstrate this decomposition for a simple graph in Figure \ref{fig:hodgeDecompAndDeformations}.

As clearly shown in the image, the specific heat from Magnetic Laplacian\cite{chaos2020} provides informative results for some components, while the Dilation Laplacian does not, and vice versa. This highlights the need for varied approaches to fully comprehend the structure and properties of a directed graph.

We aim to substantiate our claim by investigating the influence of the components, derived from the Helmholtz-Hodge decomposition, on methods based on Deformed Laplacians. We will utilize the proposed mapping of images to directed graphs, as defined in Appendix \ref{sec:img2digraphGrad}. Although we are not primarily targeting potential practical applications in image processing, our findings suggest that using this mapping in tandem with the deformation formalism may offer solutions to segmentation challenges that the combinatorial Laplacian cannot address. Our main objective is to evaluate how this decomposition influences the outcomes attained by directed graphs and magnetic Laplacians, thereby providing an  experimental method for identifying possible limitations or deficiencies in magnetic Laplacians.

\begin{figure}
	\centering
	\includegraphics[width=1\columnwidth]{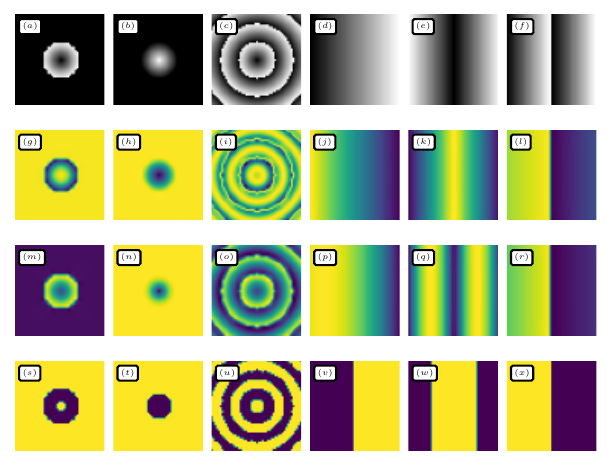}
	\caption{
		Gradient approach for image segmentation using $q=1/10$ and $\eta =0.5$. The first row represents the six intensity fields. The second and third rows depict the frustration of the $x$ axis ($\cos \Phi$) and the $y$ axis ($\sin \Phi$) respectively. The fourth row displays the absolute value of the first eigenvector of the magnetic Laplacian, while the final row illustrates the result of the K-means algorithm for $2$ clusters.}
	\label{fig:imgSegGrad}
\end{figure}

Figure \ref{fig:imgSegGrad} demonstrates the result of the gradient approach for image segmentation using $q=1/10$ and $\eta =0.5$. Interestingly, the magnetic Laplacian-based segmentation approach is capable of grouping even non-adjacent pixels. Although these results are not directly related to image processing applications, they provide valuable insights into the behavior of the graph when it loses rotational and/or gradient components. To deepen this investigation, we propose an examination of the behavior of this segmentation method within the components of the Helmholtz-Hodge decomposition, paralleling the motivation that informed the results shown in Figure \ref{fig:hodgeDecompAndDeformations}.

\begin{figure}
	\centering
	\includegraphics[width=1\columnwidth]{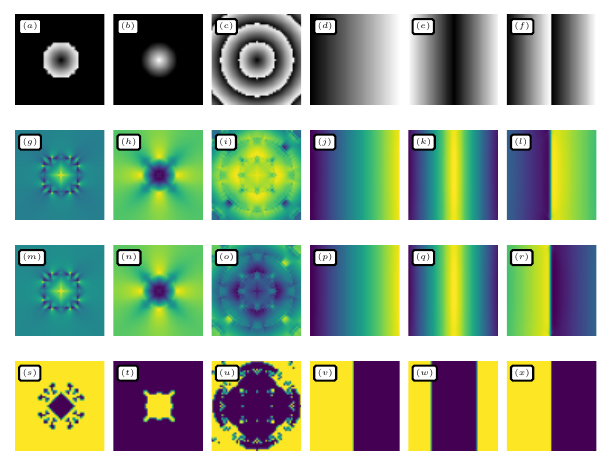}
	\caption{Gradient approach for image segmentation using $q=1/10$ and $\eta =0.5$ applied to the rotational component of the directed graph associated with each image. The first row presents the six intensity fields. The second and third rows represent the $x$-axis frustration ($\cos \Phi$) and the $y$-axis frustration ($\sin \Phi$), respectively. The fourth row illustrates the absolute value of the first eigenvector of the magnetic Laplacian. The last row displays the result of the K-means algorithm for $2$ clusters. The results in (d), (e), and (f) remain equivalent to those obtained in figure \ref{fig:imgSegGrad}.}
	\label{fig:imgSegRot}
\end{figure}

The impact of applying the magnetic Laplacian segmentation method to the rotational components is depicted in Figure \ref{fig:imgSegRot}. Effective segmentation results observed for images (a), (b), and (c) were lost. Recently, \cite{edgedir2023} provided empirical evidence that the MagNet architecture (a Graph Neural Network via magnetic Laplacians) yields less optimal results than direction-ignoring GCN architectures for specific datasets such as SQUIRREL and CHAMELEON as per \cite{squirrelPaper}. We advocate for a more thorough investigation into the weight distribution across the rotational and gradient components, as well as the sources of frustrations within these graphs. Such an exploration could formally elucidate the circumstances under which the magnetic approach falls short, mirroring its failure in some of the images presented in Figure \ref{fig:imgSegRot}.

\begin{figure}[!htb]
	\centering
	\includegraphics[width=1\columnwidth]
	{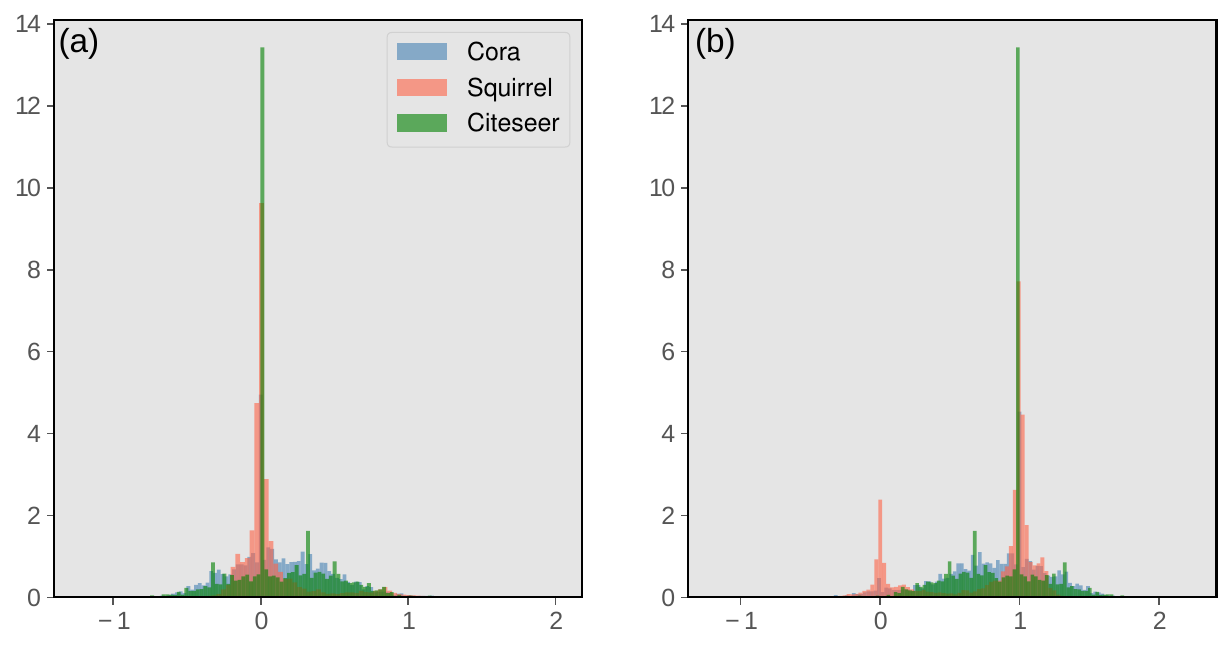}
	\caption{
		Histograms of the edge weights for the circular components (a) and the gradient components (b) for the Cora, Squirrel, and Citeseer networks. The  discrepancy in the behavior of the gradient component in the Squirrel network stands out, encouraging further studies that could aid in understanding the behavior of GCNs in specific directed networks.
	}
	\label{fig:hodgeSquirrel}
\end{figure}

To stimulate motivation for future investigations in this field, we performed numerical calculations and present in Figure \ref{fig:hodgeSquirrel} the histograms of the edge weights for the circular components (a) and the gradient components (b) for the Cora, Squirrel, and Citeseer networks. We observe a discrepancy in the behavior of the gradient component for the Squirrel network. The nature of this discrepancy, which needs to be understood and explained, emerges as a significant driver for future research aimed at understanding directed network datasets. Thus, the importance of an in-depth study in this field is highlighted, in order to clarify the emerging peculiarities and their possible consequences for the efficiency of different architectures of directed graph neural networks.

Our discussion of adjacency operators, as well as magnetic and dilation Laplacians, has revealed an inherent dilemma. These methods and measures, despite their novelty, may not be apt for all scenarios. There are cases where they fail to capture the complexity and nuances inherent in the directed networks. This poses a challenge that needs addressing.

In response to this, we propose in this paper the concept of effective graphs as
discussed in section~\ref{sec:methEffectiveMatrix}. These graphs have matrices
that are capable of incorporating information from group  deformation, and even
combinations thereof. This makes them a powerful tool for associating real,
symmetric matrices with directed (un)signed graphs. The introduction of
effective matrices opens a new path in our exploration of graphs, as we will
elucidate in the following sections through various results applying this
innovative formalism.

\subsection{Effective Graphs: Capturing New Information in the Analysis of Directed Networks}

\begin{figure*}
	\centering
	\includegraphics[width=0.9\textwidth]{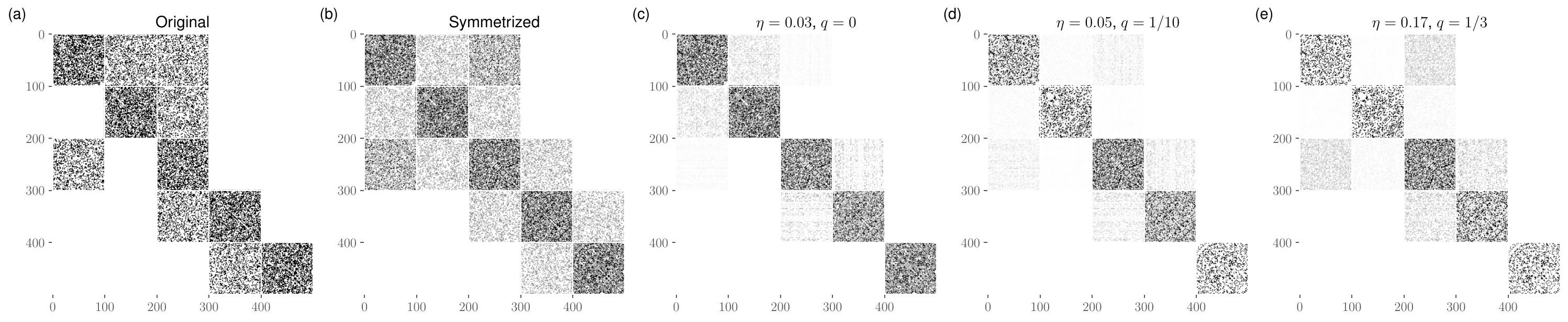}
	\caption{Visualization of the adjacency matrix of a directed graph and its symmetrical and effective representations. The original matrix is shown in (a), the symmetrized version in (b), and the effective versions, derived from equation \ref{eq:effWeight}, in (c), (d), and (e). The values of frustration ($\eta$) and charge for magnetic deformation ($q$) are indicated in the respective images. For all effective matrices, the dilation parameter is $0.1/(|V|-1)$, where $|V|=500$.}
	\label{fig:adjDecomp}
\end{figure*}

Proceeding with our discussion, Figure \ref{fig:adjDecomp} illustrates the construction of three effective adjacency matrix, each with different charge values, while holding the dilation parameter constant. In (a), we display the adjacency matrix of a directed graph, obtained from a block model. The symmetrized matrix, $(\mathbf A + \mathbf A^T)$, is shown in (b). Effective adjacency matrices, derived from equation \ref{eq:effWeight}, are illustrated in (c), (d), and (e). The frustration ($\eta$) and magnetic deformation charge ($q$) values are indicated in the titles of the respective images. For all effective matrices, the dilation parameter is set as $0.1/(|V|-1)$, where $|V|=500$.

As previously mentioned, effective graphs allow the application of all undirected graph measures to directed graphs. For example, here we explore the  \emph{betweenness centrality} measure to both the symmetrized and effective graphs, both derived from the same directed graph used to produce Figure \ref{fig:adjDecomp}. Betweenness is a widely utilized measure due to its interpretation as an indicator of information flow in a complex network\cite{newmanBC}.

The measure for a vertex $v$ can be calculated using the following equation:

\begin{align}
	B(v) = \sum\limits_{i,j} \frac{\sigma_{ij}(v)}{\sigma_{ij}},
\end{align}
where $\sigma_{ij}$ represents the number of shortest paths connecting vertices $i$ and $j$, and $\sigma_{ij}(v)$ is the number of these shortest paths passing through $v$.

It is important to note that betweenness is ill-defined for directed graphs, as the existence of a path is not necessarily guaranteed, even in a weakly connected graph. However, this does not apply to the effective graphs defined in this work.

To better understand the relations between the betweenness values of the symmetrized and effective graphs, we decided to perform an embedding of the original graph using the same approach proposed in\cite{tab2graph}. This embedding uses the phases of the magnetic Laplacian's eigenvectors and the solution to the dilation problem (absolute value of the first eigenvector), as follows:

\begin{align}
	\mathcal C_{q, g}^{(l)}:
	V&\mapsto S^1\times \mathbb R^+
	\nonumber\\
	\mathcal C_{q, g}^{(l)}(u)&=
	(
	(\angle v_q^{(l)})(u), |v_g^{(1)}(u)|
	) \\
	&= (\theta_l (u), s(u)),\nonumber
\end{align}
where $v_q^{(l)}$ and $v_g^{(0)}$ are the $l$-th eigenvector of the magnetic Laplacian with charge $g$ and the first eigenvector of the dilation Laplacian with parameter $g$, respectively.

To adjust the radial coordinate so that the vertex with the highest rank is at $0$ we perform the following mapping:
\begin{align}
	\mathcal C (u) =
	\left(
	1-\frac{s(u)-s_{\min}}{s_{\max}-s_{\min}}
	\right)
	(\cos(\theta_2(u)), \sin(\theta_2(u))),
	\label{eq:cyn2DMap}
\end{align}
where $s_{\max}=\max\limits_{u\in V}s(u)$ and $s_{\min}=\min\limits_{u\in V}s(u)$.

\begin{figure}[!htb]
	\centering
	\includegraphics[width=0.9\columnwidth]
	{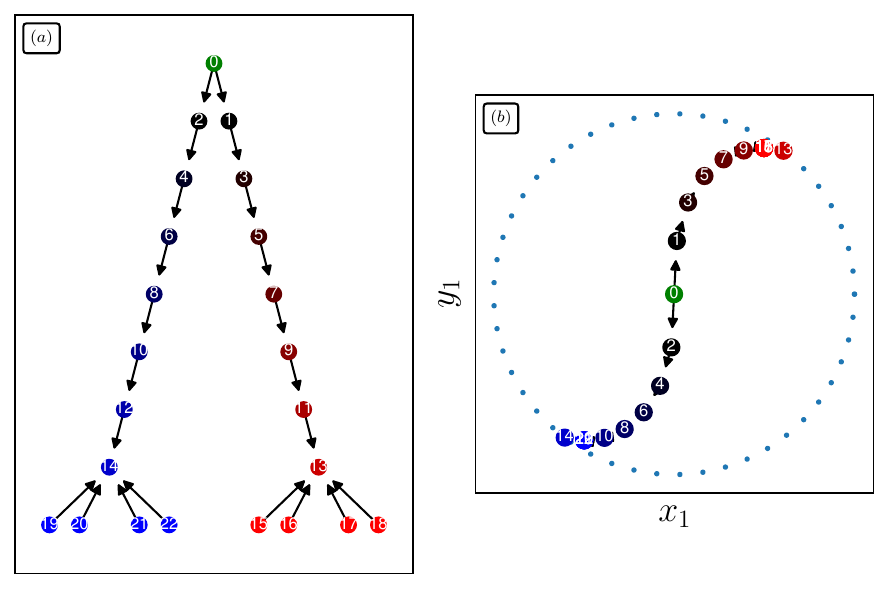}
	\caption{In (a), we present a structured directed graph with colored nodes with manually defined positions. In (b), the result of the embedding is displayed, with the phases of the first eigenvector of the Magnetic Laplacian ($q=1/3$) serving as polar coordinates and the radial coordinate determined by the ranking of the Dilation Laplacian ($g=3/10$). We perform a transformation to position the highest rank vertex at the center. Notably, high hierarchy nodes are closer to the center of the circle and the directed graph arms are immersed to reflect the original graph symmetry. This simple example serves to illustrate how applying an embedding using the solutions of frustration problems (GSP), considering both the hierarchy and cycles in a graph, can be a useful tool in analyzing directed structures.}
	\label{fig:triangleDiGraph}
\end{figure}

To illustrate how this embedding strategy could improve our understanding of a set of attributes, we consider Fig.\ref{fig:triangleDiGraph}. Sub-figure (a) presents a structured directed graph, with node colors used for visual inspection and manually adjusted positions. Sub-figure (b) illustrates the results of the mapping to the space that combines the solutions to the magnetic frustration and dilation problem.

In this simple example, the embedding shown in Fig.\ref{fig:triangleDiGraph} clearly reveals the hierarchy and symmetrical relationships inherent in the directed graph. With the high ranked vertex closer to the center and the preservation of graph ``arm'' symmetry, this example provides an intuitive visual representation of what we expected when applying

\begin{figure}
	\centering
	\includegraphics[width=1\columnwidth,keepaspectratio]{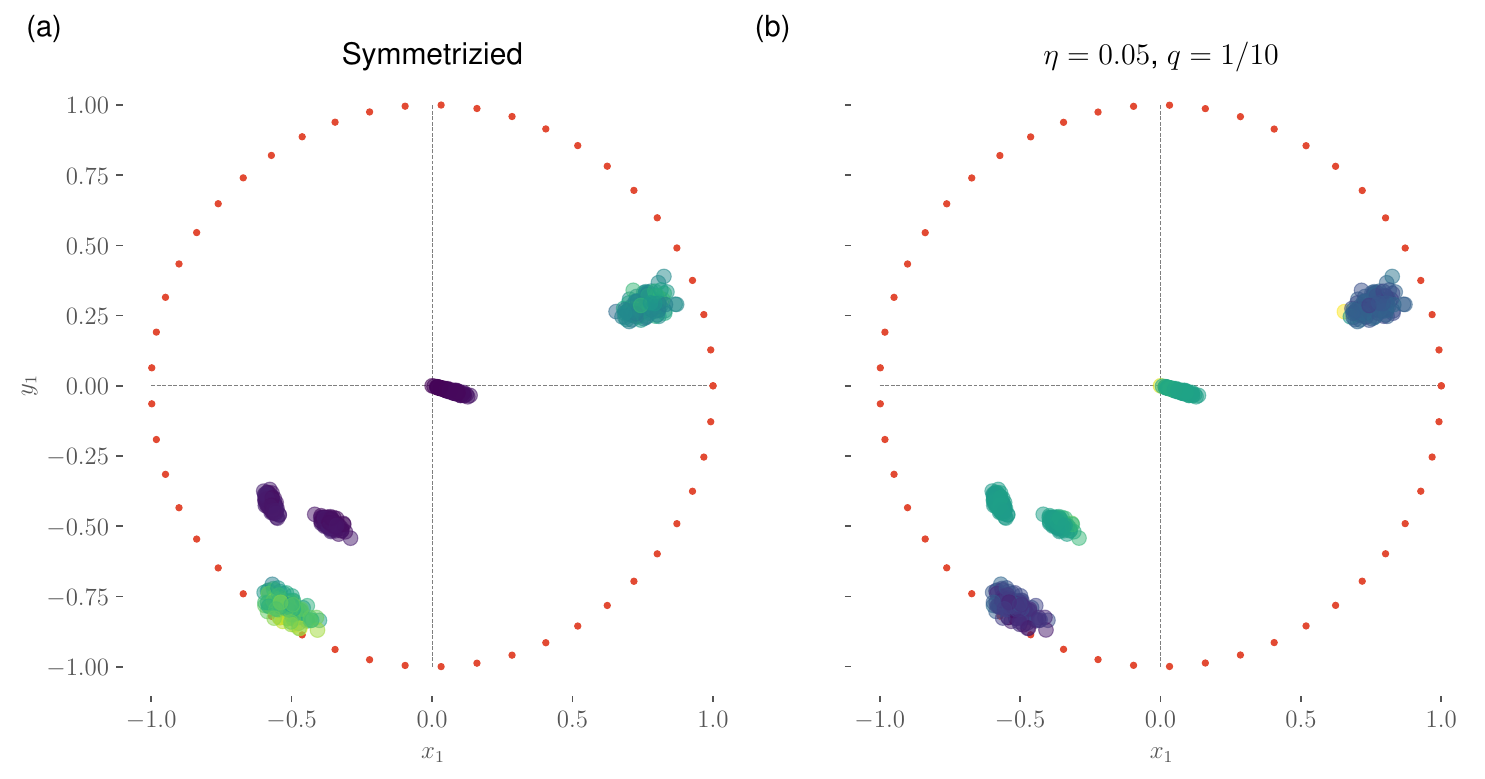}
	\caption{Embedding of a directed graph using the solutions to the magnetic Laplacian and the dilation problem, as described in the main text. The color of each node corresponds to the value of its betweenness centrality in the  symmetrized and effective  graph. The node positioning was achieved through mapping to the space that combines the solutions to the magnetic frustration and dilation problem, as explained previously. }
	 \label{fig:centralidadeEffetiva}
\end{figure}

In Figure \ref{fig:centralidadeEffetiva}, we present the embedding of the directed graph used to obtain the result shown in Figure  \ref{fig:adjDecomp}
using the methodology previously described. Each node is colored according to its betweenness centrality in the original and effective graph. It becomes evident that the measures taken from the symmetrized graph  do not reflect the complexity of the original directed graph, while the effective graph offers a better understanding of the original structure.

Although the symmetrization process simplifies the network analysis, this approach disregards the directionality of the connections. Conversely, the effective graph captures more features from the original directed graph.

In conclusion, the effective graph  allows the use of traditional network measures, which are designed for undirected graphs, opening new paths for the analysis and understanding of complex networks.

\subsection{Effective Coarse-Graining and Renormalization Groups}

The concept of Renormalization Group (RG) is a fundamental pillar in universality theory and phase transitions\cite{Anderson_1972,Goldenfeld_1992}. However, their application in complex networks has been challenging due to the heterogeneity of these networks.

Recently, the Laplacian Renormalization Group (RGL) was defined \cite{rgNature}, which introduced a new perspective based on diffusion in a complex network. The RGL is capable of identifying appropriate space-time scales in heterogeneous networks, representing an advance over previous methods, such as those based on neighborhoods. Recently, it has been shown how this technique can enhance the quality of Graph Neural Networks \cite{rgGCN}. A generalization for hypergraphs (high-order networks) was proposed in \cite{rgHighOrder}, while the work of  \cite{rgBrasil} formalized the RGL approach using the formalism of field theory, widely used in quantum field theory. However, all these RG methods are not applicable in more general contexts, such as directed and/or signed graphs.

Given these limitations, there is an imperative need for the evolution of renormalization group methodologies that can work with directed and/or signed graphs. Such an extension will enable more comprehensive analyses of a broad range of complex systems and datasets, as many real-world networks are indeed directed. In this paper, we will discuss how RG can be leveraged using the effective graph formalism. Moreover,  this novel approach can be utilized to assert that the effective graph formalism yields more than mere simple symmetrization.

The procedure for using a renormalization process in effective graphs is described in the box below.

\begin{tcolorbox}
	\begin{flushleft}

		To implement a renormalization process on effective graphs, the following elements are required: a directed graph, $G$,
		a charge value $q$, a dilation parameter $g$, a scale parameter $\beta$, a strategy
		$\mathcal S$ for solving the associated frustration problem, a sparsification method (edge removal)
		$\mathcal E$, and finally a selection process The steps are as follows:
		\begin{itemize}
			\item construct the deformed Laplacian using parameters $g$ and $q$;
			\item compute the frustration using equation \ref{eqSync} and a strategy $\mathcal S$;
			\item calculate the effective weight for each edge using equation
			\ref{eq:effWeight}, which gives an undirected graph $G_U$;

			\item sparsify $G_U$ using a technique $\mathcal E$;
			\item contract $|V|//2$  pairs of vertices from	 $\mathcal G$  applied on $G_U$ into $|V|//2$ super-vertices.
			\item create a directed graph $\tilde G$ with $|V|//2$ vertices. An edge between two vertices is established if there was at least one edge between the original vertices.
		\end{itemize}
	\end{flushleft}
\end{tcolorbox}

The RGEG (Renormalized Graph Effective Graph), as outlined above, is highly flexible.
Here, for simplicity, for $\mathcal E$, we chose the disparity filter that we used in \cite{tab2graph} and which was implemented in the open-source library EdgeEraser \cite{messias_2022}. For $\mathcal{S}$, the process is described in the box below.

\begin{tcolorbox}
	\begin{flushleft}
		\begin{itemize}
			\item We construct a normalized magnetic Laplacian with charge $q$;
			\item We extract the phases from the first eigenvector of the magnetic Laplacian and store them $\mathbf \theta_1$;
\item we construct a dilation Laplacian with parameter $g$;
\item we obtain an approximate solution for the dilation frustration problem, represented by the vector formed by the norm of each component of the first eigenvector of the dilation Laplacian, $\mathbf s$;
		\item we construct the solution vector whose component $u$ is given by:
		$(e^{\ui q \theta_1}s)(u)$.
		\end{itemize}
	\end{flushleft}
\end{tcolorbox}

To perform the coarse-graining in the effective undirected graph we chose the pseudoinverse  of the combinatorial Laplacian, as discussed in \cite{rgBrasil} . Thus, the
$|V|//2$ pairs of vertices to be contracted represent a notion of correlation, meaning the super-vertices will be formed by the most correlated pairs of vertices.

As a case study to compare the coarse-graining process in both the symmetrized and effective versions, we sampled a graph obtained from a block model, keeping the same parameters used in the graph generation to obtain the results of\cite{arxiv2018}. That is, a graph with a cyclic structure with $N_f=3$ blocks, where each undirected block has a connection probability of $p_c = 50\%$ and directed connections between blocks with a probability of $p_d = 70\%$.

\begin{figure}
	\centering
	\includegraphics[width=1\columnwidth,keepaspectratio]{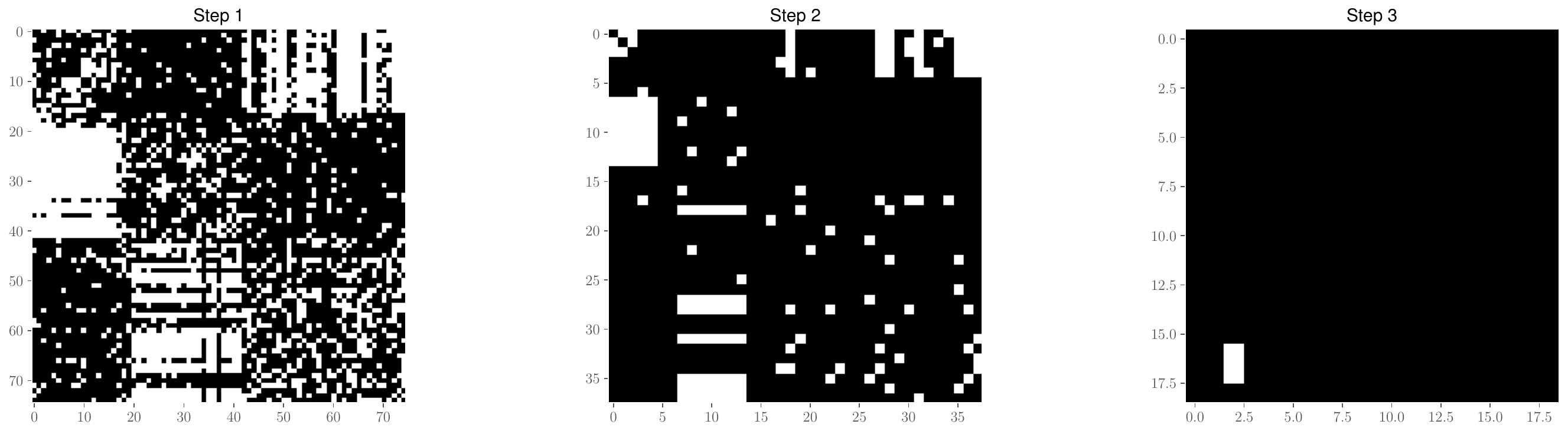}
	\caption{
		Evolution of the adjacency matrix obtained through the RG process for the Symmetrized Graph derived from a directed graph with 3 blocks. Note the homogenization process indicating that the RG approach does not adequately apply to the directed graph using the symmetric adjacency matrix.
	}
	\label{fig:rgegSym}
\end{figure}

Figure \ref{fig:rgegSym} shows the RG (Renormalization Group) result, considering only the symmetrized graph, that is, it is equivalent to setting $q=g=0$. By the third step, all information about directionality was lost as the graph became homogenized into a fully connected network. Therefore, this demonstrates that the RG approach is not useful when the symmetrized version of the graph is used.

\begin{figure}
	\centering
	\includegraphics[width=1\columnwidth,keepaspectratio]{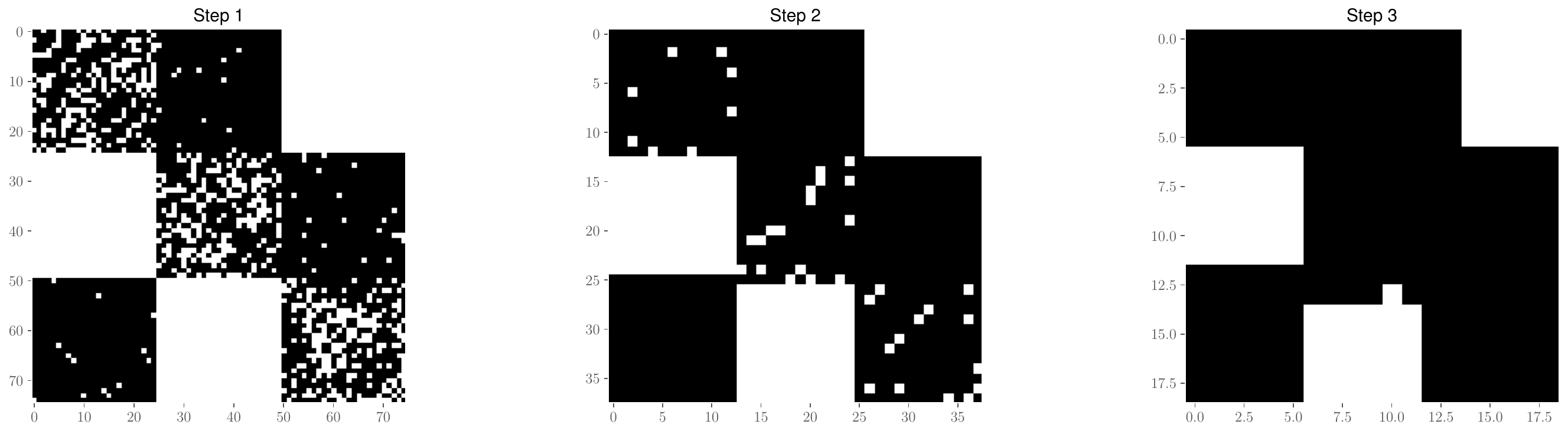}
	\caption{Evolution of the adjacency matrix obtained through the RGEG process for the Effective Graph. Four RGEG steps applied using $q=1/10$ and $g=0$ n a directed graph with three blocks in a cycle. The RGEG process reinforces the cyclic structure and preserves the direction. }
	\label{fig:rgegEffective}
\end{figure}

Figure \ref{fig:rgegEffective} presents the RGEG (Effective Graph Renormalization Group) result by using the parameters $q=1/10$ and $g=0$. It is observed that the RGEG process strengthens the block structure and the directionality of the cycles in the model. In the fourth step, we have a directed graph that could have been obtained with parameters $N_f=3$, $p_d=100\%$, and $p_c=100\%$. Therefore, these results are a promising indication that we can explore RGEG in other contexts.

We will analyze the application of the coarse-graining method on a real dataset, specifically the blogosphere of the 2004 U.S. elections \cite{polblogs2005}. This historical event highlighted the growing influence of the blogosphere on the electoral process. Each vertex in this network represents a blog, with its political stance (Democrat or Republican) manually annotated. The edges represent hyperlinks between blogs.

The blogosphere during the 2004 elections was pivotal, as it marked a turning point in how politics are conducted and communicated. Given its historical significance and the quality of this dataset, it continues to be a valuable resource for evaluating graph processing techniques.

\begin{figure*}
	\centering
	\includegraphics[width=1\textwidth]{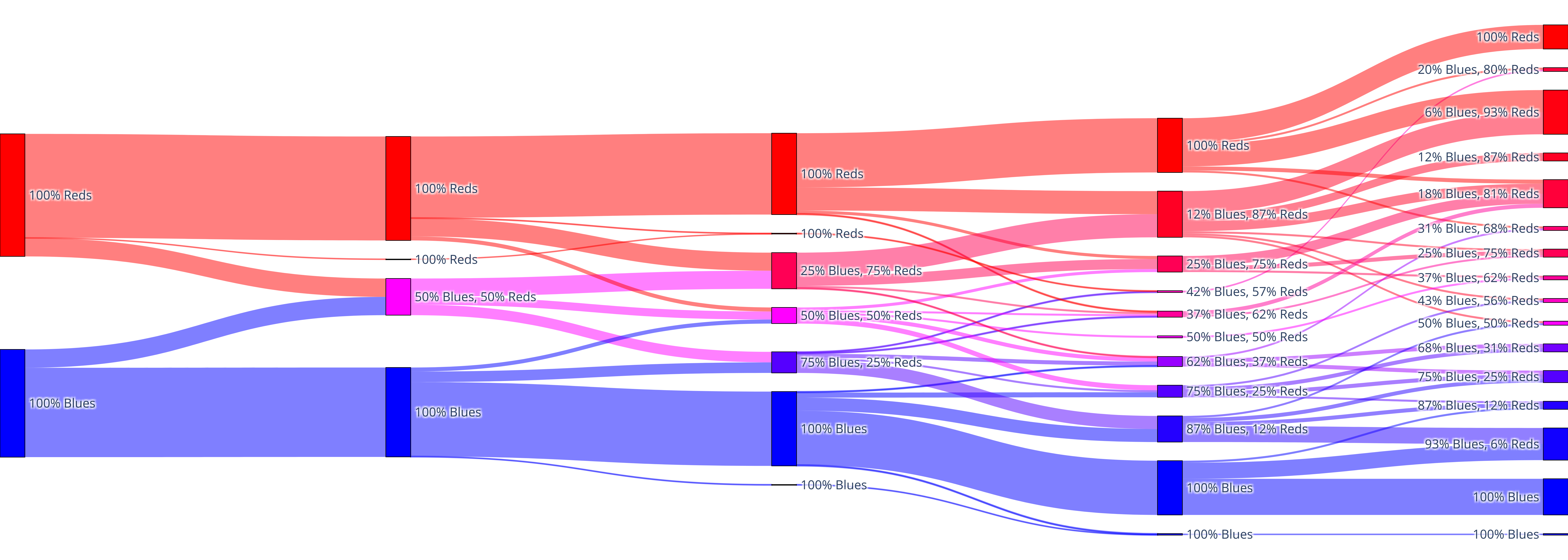}
	\caption{Application of the RGEG flow on the American blogosphere of 2004.\cite{polblogs2005} In the initial state, each vertex, corresponding to a blog, is colored blue to represent Democrats and red to represent Republicans. Upon applying the RGEG, it is observed that the vertices tend to form clusters with similar colors, indicating minimal interaction between blogs of different political orientations. This result highlights the efficacy of the RGEG method in preserving the inherent structure of the network and the separation between different political inclinations.
	}
	\label{fig:polBlogsFlow}
\end{figure*}

Figure \ref{fig:polBlogsFlow} shows the clustering flow resulting from the application of the coarse-graining in effective matrices on the network (from left to right), where vertices are color-coded to represent Democrats (blue) and Republicans (red). In the initial step, before the application of RGEG, each vertex is in an isolated state. The analysis reveals that the coarse-graining by RGEG, when applied to the directed graph of the blogosphere, results in minimal interaction between blogs of different political leanings. This outcome, in turn, reinforces the validity and effectiveness of our RGEG.

\begin{figure}
	\centering
	\includegraphics[width=1\columnwidth,keepaspectratio]{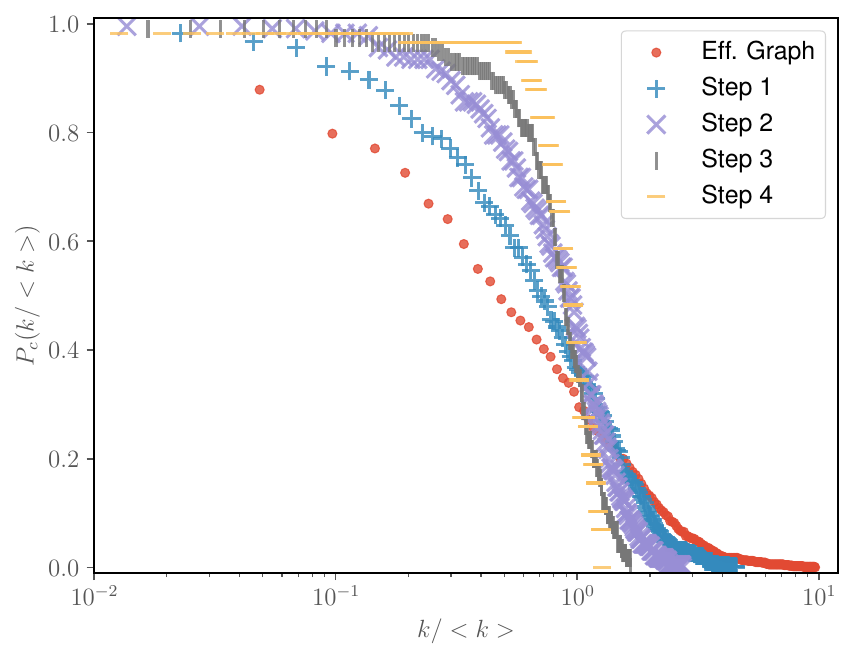}
	\caption{
		Cumulative degree distribution for the effective graph and for four steps of coarse-graining. The coarse-graining method selected was the same as the one applied in \cite{rgBrasil}. Therefore, similar to that study, it suffers from the lack of renormalization in the edge weights.
	}
\label{fig:ccdf}
\end{figure}

Figure~\ref{fig:ccdf} depicts the cumulative degree distribution for the effective graph and for four steps of coarse-graining. The method employed for coarse-graining is consistent with the one used in \cite{rgBrasil}. Similar to the findings in that study, this approach also encounters challenges due to the absence of renormalization in edge weights. This aspect is crucial as it highlights the limitations of the current methodology, particularly in maintaining the integrity of edge weights through the coarse-graining steps, which is essential for preserving the network's topological characteristics.

\begin{figure}
	\centering
	\includegraphics[width=1\columnwidth,keepaspectratio]{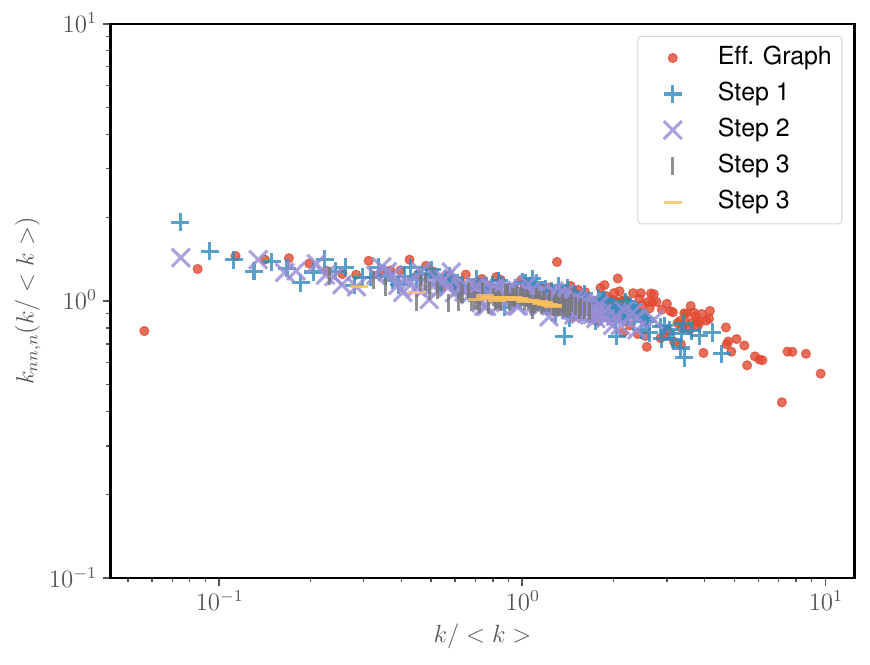}
	\caption{
		Degree correlation among neighbors vertex for the effective graph of the American blogosphere and four steps of coarse-graining. An indication of scale invariance can be observed.
	}
\label{fig:knn}
\end{figure}

Figure~\ref{fig:knn} presents the degree correlation among neighboring vertices for the effective graph of the American blogosphere and during four steps of coarse-graining. This figure provides an indication of scale invariance, a fundamental property in many real-world networks that suggests a consistent pattern of connectivity regardless of the scale at which the network is observed. This property is particularly relevant in the context of political discourse within the blogosphere, as it may reflect the inherent structure of political affiliations and discourse patterns.

Together, these figures complement the initial analysis by providing a quantitative view of the network's evolution under coarse-graining. They reveal the method's effects on network topology, including degree distribution and vertex connectivity, which are critical for understanding the underlying structure and dynamics of the blogosphere in the context of the 2004 U.S. elections.

	\section{Conclusions}
In this work, we ventured into the realm of directed and/or signed graphs,
highlighting the advent of new Laplacian operators to address their
peculiarities. While magnetic Laplacians and their variations have proven useful
in the literature, they come with the trade-off of necessitating the development
of new metrics, machine learning methods, and new code implementations. This
raises a critical question: Can we circumvent this complexity while still
gaining meaningful insights into the network structure?

Our exploration has revealed that despite the widespread adoption of magnetic
Laplacian formalism for characterizing directed networks, it does not always
capture the full spectrum of information inherent to the network's
directionality. In certain cases, we found that the magnetic approach might be
less effective than methodologies that ignore directionality, such as the
dilation Laplacian. This observation underscores the importance of diverse
methodologies to fully understand the intricacies of directed graphs.

We introduced the concept of effective graphs, which allows the application of
traditional network measures and methods, typically reserved for undirected,
unsigned graphs. This novel approach opens new avenues for the analysis and
understanding of complex networks. Our investigation into effective graphs
revealed that while symmetrization processes simplify network analysis, they
might overlook critical aspects of connection directionality. In contrast,
effective graphs manage to retain more characteristics from the original
directed graphs, offering a richer understanding of network structure.

Moreover, we delved into the applicability of the Renormalization Group concept
to effective graphs, marking a significant advancement over previous methods.
These methods have proven challenging in   directed and/or signed complex
networks. The results indicated that renormalization in effective graphs could
preserve and even reinforce block structures and directionality in graph models,
offering promising insights for broader network analysis.

To further expand upon these findings and explore new directions, we propose the
following future works:

\begin{itemize}
	\item \textbf{Application of Graph Neural Network Techniques on Effective Graphs:} Applying Graph Neural Network (GNN) models to undirected networks within datasets that are inherently directed but have been mapped onto effective graphs. This involves comparing their performance with that of magnetic Laplacian-based approaches and methodologies for undirected graphs.
	\item \textbf{Application of Effective Matrix Techniques to Undirected Graphs with Real Flows:} Extending the effective matrix approach to analyze undirected graphs that have real flows defined on their edges, which is common in mobility and transportation datasets.
	\item\textbf{Studying the effect of renormalization in effective graphs:} to potentially enhance the outcomes of GCN on directed graphs as has been proven useful in undirected datasets.
	\item\textbf{Exploration of dataset of directed graphs  using Helmohtz decomposition and deformed Laplacian methods.}
\end{itemize}

	\bibliography{refs.bib}

\begin{thebibliography}{42}%
\makeatletter
\providecommand \@ifxundefined [1]{%
 \@ifx{#1\undefined}
}%
\providecommand \@ifnum [1]{%
 \ifnum #1\expandafter \@firstoftwo
 \else \expandafter \@secondoftwo
 \fi
}%
\providecommand \@ifx [1]{%
 \ifx #1\expandafter \@firstoftwo
 \else \expandafter \@secondoftwo
 \fi
}%
\providecommand \natexlab [1]{#1}%
\providecommand \enquote  [1]{``#1''}%
\providecommand \bibnamefont  [1]{#1}%
\providecommand \bibfnamefont [1]{#1}%
\providecommand \citenamefont [1]{#1}%
\providecommand \href@noop [0]{\@secondoftwo}%
\providecommand \href [0]{\begingroup \@sanitize@url \@href}%
\providecommand \@href[1]{\@@startlink{#1}\@@href}%
\providecommand \@@href[1]{\endgroup#1\@@endlink}%
\providecommand \@sanitize@url [0]{\catcode `\\12\catcode `\$12\catcode
  `\&12\catcode `\#12\catcode `\^12\catcode `\_12\catcode `\%12\relax}%
\providecommand \@@startlink[1]{}%
\providecommand \@@endlink[0]{}%
\providecommand \url  [0]{\begingroup\@sanitize@url \@url }%
\providecommand \@url [1]{\endgroup\@href {#1}{\urlprefix }}%
\providecommand \urlprefix  [0]{URL }%
\providecommand \Eprint [0]{\href }%
\providecommand \doibase [0]{http://dx.doi.org/}%
\providecommand \selectlanguage [0]{\@gobble}%
\providecommand \bibinfo  [0]{\@secondoftwo}%
\providecommand \bibfield  [0]{\@secondoftwo}%
\providecommand \translation [1]{[#1]}%
\providecommand \BibitemOpen [0]{}%
\providecommand \bibitemStop [0]{}%
\providecommand \bibitemNoStop [0]{.\EOS\space}%
\providecommand \EOS [0]{\spacefactor3000\relax}%
\providecommand \BibitemShut  [1]{\csname bibitem#1\endcsname}%
\let\auto@bib@innerbib\@empty
\bibitem [{\citenamefont {Zheng}\ and\ \citenamefont
  {Zheng}(2017)}]{zhengSocialNetworksRich2017a}%
  \BibitemOpen
  \bibfield  {author} {\bibinfo {author} {\bibfnamefont {Quan}\ \bibnamefont
  {Zheng}}\ and\ \bibinfo {author} {\bibfnamefont {Quan}\ \bibnamefont
  {Zheng}},\ }\href@noop {} {\emph {\bibinfo {title} {Social Networks with Rich
  Edge Semantics}}},\ \bibinfo {series} {Chapman \& {{Hall}}/{{CRC}} Data
  Mining and Knowledge Discovery}\ No.~\bibinfo {number} {30}\ (\bibinfo
  {publisher} {{Taylor \& Francis, CRC Press}},\ \bibinfo {address} {{Boca
  Raton}},\ \bibinfo {year} {2017})\BibitemShut {NoStop}%
\bibitem [{\citenamefont {Harary}(1953{\natexlab{a}})}]{Harary_1953}%
  \BibitemOpen
  \bibfield  {author} {\bibinfo {author} {\bibfnamefont {Frank}\ \bibnamefont
  {Harary}},\ }\bibfield  {title} {\enquote {\bibinfo {title} {On the notion of
  balance of a signed graph.}}\ }\href {\doibase 10.1307/mmj/1028989917}
  {\bibfield  {journal} {\bibinfo  {journal} {The Michigan Mathematical
  Journal}\ }\textbf {\bibinfo {volume} {2}},\ \bibinfo {pages} {143--146}
  (\bibinfo {year} {1953}{\natexlab{a}})}\BibitemShut {NoStop}%
\bibitem [{\citenamefont
  {Aittokallio}(2006)}]{aittokallioGraphbasedMethodsAnalysing2006}%
  \BibitemOpen
  \bibfield  {author} {\bibinfo {author} {\bibfnamefont {T.}~\bibnamefont
  {Aittokallio}},\ }\bibfield  {title} {\enquote {\bibinfo {title} {Graph-based
  methods for analysing networks in cell biology},}\ }\href {\doibase
  10.1093/bib/bbl022} {\bibfield  {journal} {\bibinfo  {journal} {Briefings in
  Bioinformatics}\ }\textbf {\bibinfo {volume} {7}},\ \bibinfo {pages}
  {243--255} (\bibinfo {year} {2006})}\BibitemShut {NoStop}%
\bibitem [{\citenamefont {Zheng}\ and\ \citenamefont
  {Skillicorn}(2015)}]{zhengAnalysisCriminalSocial2015}%
  \BibitemOpen
  \bibfield  {author} {\bibinfo {author} {\bibfnamefont {Quan}\ \bibnamefont
  {Zheng}}\ and\ \bibinfo {author} {\bibfnamefont {David~B.}\ \bibnamefont
  {Skillicorn}},\ }\bibfield  {title} {\enquote {\bibinfo {title} {Analysis of
  criminal social networks with typed and directed edges},}\ }in\ \href
  {\doibase 10.1109/ISI.2015.7165930} {\emph {\bibinfo {booktitle} {2015 {{IEEE
  International Conference}} on {{Intelligence}} and {{Security Informatics}}
  ({{ISI}})}}}\ (\bibinfo  {publisher} {IEEE},\ \bibinfo {address} {Baltimore,
  MD, USA},\ \bibinfo {year} {2015})\ pp.\ \bibinfo {pages} {1--6}\BibitemShut
  {NoStop}%
\bibitem [{\citenamefont {Li}\ \emph {et~al.}(2018)\citenamefont {Li},
  \citenamefont {Yuan}, \citenamefont {Wu},\ and\ \citenamefont
  {Lu}}]{Li_Yuan_Wu_Lu_2018}%
  \BibitemOpen
  \bibfield  {author} {\bibinfo {author} {\bibfnamefont {Yuemeng}\ \bibnamefont
  {Li}}, \bibinfo {author} {\bibfnamefont {Shuhan}\ \bibnamefont {Yuan}},
  \bibinfo {author} {\bibfnamefont {Xintao}\ \bibnamefont {Wu}}, \ and\
  \bibinfo {author} {\bibfnamefont {Aidong}\ \bibnamefont {Lu}},\ }\bibfield
  {title} {\enquote {\bibinfo {title} {On spectral analysis of directed signed
  graphs},}\ }\href {\doibase 10.1007/s41060-018-0143-9} {\bibfield  {journal}
  {\bibinfo  {journal} {International Journal of Data Science and Analytics}\
  }\textbf {\bibinfo {volume} {6}},\ \bibinfo {pages} {147--162} (\bibinfo
  {year} {2018})}\BibitemShut {NoStop}%
\bibitem [{\citenamefont {Fanuel}\ and\ \citenamefont
  {Suykens}(2019)}]{fanuelDeformedLaplaciansSpectral2019}%
  \BibitemOpen
  \bibfield  {author} {\bibinfo {author} {\bibfnamefont {M.}~\bibnamefont
  {Fanuel}}\ and\ \bibinfo {author} {\bibfnamefont {J.A.K.}\ \bibnamefont
  {Suykens}},\ }\bibfield  {title} {\enquote {\bibinfo {title} {Deformed
  laplacians and spectral ranking in directed networks},}\ }\href {\doibase
  10.1016/j.acha.2017.09.002} {\bibfield  {journal} {\bibinfo  {journal}
  {Applied and Computational Harmonic Analysis}\ }\textbf {\bibinfo {volume}
  {47}},\ \bibinfo {pages} {397--422} (\bibinfo {year} {2019})}\BibitemShut
  {NoStop}%
\bibitem [{\citenamefont {Shahriari}\ and\ \citenamefont
  {Jalili}(2014)}]{shahriariRankingNodesSigned2014}%
  \BibitemOpen
  \bibfield  {author} {\bibinfo {author} {\bibfnamefont {Moshen}\ \bibnamefont
  {Shahriari}}\ and\ \bibinfo {author} {\bibfnamefont {Mahdi}\ \bibnamefont
  {Jalili}},\ }\bibfield  {title} {\enquote {\bibinfo {title} {Ranking nodes in
  signed social networks},}\ }\href {\doibase 10.1007/s13278-014-0172-x}
  {\bibfield  {journal} {\bibinfo  {journal} {Social Network Analysis and
  Mining}\ }\textbf {\bibinfo {volume} {4}},\ \bibinfo {pages} {172} (\bibinfo
  {year} {2014})}\BibitemShut {NoStop}%
\bibitem [{\citenamefont {Gemici}\ and\ \citenamefont
  {Vashevko}(2018)}]{gemiciVisualizingHierarchicalSocial2018}%
  \BibitemOpen
  \bibfield  {author} {\bibinfo {author} {\bibfnamefont {Kurtulu{\c s}}\
  \bibnamefont {Gemici}}\ and\ \bibinfo {author} {\bibfnamefont {Anthony}\
  \bibnamefont {Vashevko}},\ }\bibfield  {title} {\enquote {\bibinfo {title}
  {Visualizing hierarchical social networks},}\ }\href {\doibase
  10.1177/2378023118772982} {\bibfield  {journal} {\bibinfo  {journal} {Socius:
  Sociological Research for a Dynamic World}\ }\textbf {\bibinfo {volume}
  {4}},\ \bibinfo {pages} {237802311877298} (\bibinfo {year}
  {2018})}\BibitemShut {NoStop}%
\bibitem [{\citenamefont {De~Bacco}\ \emph {et~al.}(2018)\citenamefont
  {De~Bacco}, \citenamefont {Larremore},\ and\ \citenamefont
  {Moore}}]{debaccoPhysicalModelEfficient2018}%
  \BibitemOpen
  \bibfield  {author} {\bibinfo {author} {\bibfnamefont {Caterina}\
  \bibnamefont {De~Bacco}}, \bibinfo {author} {\bibfnamefont {Daniel~B.}\
  \bibnamefont {Larremore}}, \ and\ \bibinfo {author} {\bibfnamefont
  {Cristopher}\ \bibnamefont {Moore}},\ }\bibfield  {title} {\enquote {\bibinfo
  {title} {A physical model for efficient ranking in networks},}\ }\href
  {\doibase 10.1126/sciadv.aar8260} {\bibfield  {journal} {\bibinfo  {journal}
  {Science Advances}\ }\textbf {\bibinfo {volume} {4}},\ \bibinfo {pages}
  {eaar8260} (\bibinfo {year} {2018})}\BibitemShut {NoStop}%
\bibitem [{\citenamefont {Jiang}\ \emph {et~al.}(2011)\citenamefont {Jiang},
  \citenamefont {Lim}, \citenamefont {Yao},\ and\ \citenamefont
  {Ye}}]{jiangStatisticalRankingCombinatorial2011}%
  \BibitemOpen
  \bibfield  {author} {\bibinfo {author} {\bibfnamefont {Xiaoye}\ \bibnamefont
  {Jiang}}, \bibinfo {author} {\bibfnamefont {Lek-Heng}\ \bibnamefont {Lim}},
  \bibinfo {author} {\bibfnamefont {Yuan}\ \bibnamefont {Yao}}, \ and\ \bibinfo
  {author} {\bibfnamefont {Yinyu}\ \bibnamefont {Ye}},\ }\bibfield  {title}
  {\enquote {\bibinfo {title} {Statistical ranking and combinatorial hodge
  theory},}\ }\href {\doibase 10.1007/s10107-010-0419-x} {\bibfield  {journal}
  {\bibinfo  {journal} {Mathematical Programming}\ }\textbf {\bibinfo {volume}
  {127}},\ \bibinfo {pages} {203--244} (\bibinfo {year} {2011})}\BibitemShut
  {NoStop}%
\bibitem [{\citenamefont {Pereira}\ \emph {et~al.}(2019)\citenamefont
  {Pereira}, \citenamefont {Lunardi}, \citenamefont {Calçada},\ and\
  \citenamefont {Viviane~A}}]{hodge2019br}%
  \BibitemOpen
  \bibfield  {author} {\bibinfo {author} {\bibfnamefont {Ana~L}\ \bibnamefont
  {Pereira}}, \bibinfo {author} {\bibfnamefont {José~T}\ \bibnamefont
  {Lunardi}}, \bibinfo {author} {\bibfnamefont {Marcos}\ \bibnamefont
  {Calçada}}, \ and\ \bibinfo {author} {\bibfnamefont {Bagio}\ \bibnamefont
  {Viviane~A}},\ }\bibfield  {title} {\enquote {\bibinfo {title} {Hodgerank as
  a quantitative tool in social representations theory},}\ }\href {\doibase
  10.1088/1742-6596/1391/1/012114} {\bibfield  {journal} {\bibinfo  {journal}
  {Journal of Physics: Conference Series}\ }\textbf {\bibinfo {volume}
  {1391}},\ \bibinfo {pages} {012114} (\bibinfo {year} {2019})}\BibitemShut
  {NoStop}%
\bibitem [{\citenamefont {Oliveira}\ \emph {et~al.}(2024)\citenamefont
  {Oliveira}, \citenamefont {Lunardi}, \citenamefont {Calçada}, \citenamefont
  {Pereira}, \citenamefont {De~Jesuz},\ and\ \citenamefont
  {Costa}}]{hodge2024br}%
  \BibitemOpen
  \bibfield  {author} {\bibinfo {author} {\bibfnamefont {Luna R.~N.}\
  \bibnamefont {Oliveira}}, \bibinfo {author} {\bibfnamefont {José~T.}\
  \bibnamefont {Lunardi}}, \bibinfo {author} {\bibfnamefont {Marcos}\
  \bibnamefont {Calçada}}, \bibinfo {author} {\bibfnamefont {Ana~L.}\
  \bibnamefont {Pereira}}, \bibinfo {author} {\bibfnamefont {Danilo A.~F.}\
  \bibnamefont {De~Jesuz}}, \ and\ \bibinfo {author} {\bibfnamefont {Cristina}\
  \bibnamefont {Costa}},\ }\bibfield  {title} {\enquote {\bibinfo {title}
  {Hodgerank as a new tool to explore the structure of a social
  representation},}\ }\href {\doibase 10.3389/fphy.2024.1333727} {\bibfield
  {journal} {\bibinfo  {journal} {Frontiers in Physics}\ }\textbf {\bibinfo
  {volume} {12}},\ \bibinfo {pages} {1333727} (\bibinfo {year}
  {2024})}\BibitemShut {NoStop}%
\bibitem [{\citenamefont {Gebhart}\ \emph {et~al.}(2021)\citenamefont
  {Gebhart}, \citenamefont {Fu},\ and\ \citenamefont
  {Funk}}]{hodgeGowithTheFlow}%
  \BibitemOpen
  \bibfield  {author} {\bibinfo {author} {\bibfnamefont {Thomas}\ \bibnamefont
  {Gebhart}}, \bibinfo {author} {\bibfnamefont {Xiaojun}\ \bibnamefont {Fu}}, \
  and\ \bibinfo {author} {\bibfnamefont {Russell~J.}\ \bibnamefont {Funk}},\
  }\bibfield  {title} {\enquote {\bibinfo {title} {Go with the flow? a
  large-scale analysis of health care delivery networks in the united states
  using hodge theory},}\ }in\ \href {\doibase
  10.1109/BigData52589.2021.9671805} {\emph {\bibinfo {booktitle} {2021 IEEE
  International Conference on Big Data (Big Data)}}}\ (\bibinfo {year} {2021})\
  pp.\ \bibinfo {pages} {3812--3823}\BibitemShut {NoStop}%
\bibitem [{\citenamefont {Fujiki}\ and\ \citenamefont
  {Haruna}(2015)}]{hodgeDynamics}%
  \BibitemOpen
  \bibfield  {author} {\bibinfo {author} {\bibfnamefont {Yuuya}\ \bibnamefont
  {Fujiki}}\ and\ \bibinfo {author} {\bibfnamefont {Taichi}\ \bibnamefont
  {Haruna}},\ }\bibfield  {title} {\enquote {\bibinfo {title} {Hodge
  decomposition of information flow on complex networks},}\ }in\ \href
  {\doibase 10.4108/icst.bict.2014.257876} {\emph {\bibinfo {booktitle}
  {Proceedings of the 8th International Conference on Bio-inspired Information
  and Communications Technologies (formerly BIONETICS)}}}\ (\bibinfo
  {publisher} {ACM},\ \bibinfo {address} {Boston, United States},\ \bibinfo
  {year} {2015})\BibitemShut {NoStop}%
\bibitem [{\citenamefont {Fujiwara}\ and\ \citenamefont
  {Islam}(2020)}]{hodgebitcoin}%
  \BibitemOpen
  \bibfield  {author} {\bibinfo {author} {\bibfnamefont {Yoshi}\ \bibnamefont
  {Fujiwara}}\ and\ \bibinfo {author} {\bibfnamefont {Rubaiyat}\ \bibnamefont
  {Islam}},\ }\enquote {\bibinfo {title} {Hodge decomposition of bitcoin money
  flow},}\ in\ \href {\doibase 10.1007/978-981-15-4498-9_7} {\emph {\bibinfo
  {booktitle} {Advanced Studies of Financial Technologies and Cryptocurrency
  Markets}}},\ \bibinfo {editor} {edited by\ \bibinfo {editor} {\bibfnamefont
  {Lukáš}\ \bibnamefont {Pichl}}, \bibinfo {editor} {\bibfnamefont
  {Cheoljun}\ \bibnamefont {Eom}}, \bibinfo {editor} {\bibfnamefont {Enrico}\
  \bibnamefont {Scalas}}, \ and\ \bibinfo {editor} {\bibfnamefont {Taisei}\
  \bibnamefont {Kaizoji}}}\ (\bibinfo  {publisher} {Springer Singapore},\
  \bibinfo {address} {Singapore},\ \bibinfo {year} {2020})\ p.\ \bibinfo
  {pages} {117–137}\BibitemShut {NoStop}%
\bibitem [{\citenamefont
  {Harary}(1953{\natexlab{b}})}]{hararyNotionBalanceSigned1953}%
  \BibitemOpen
  \bibfield  {author} {\bibinfo {author} {\bibfnamefont {Frank}\ \bibnamefont
  {Harary}},\ }\bibfield  {title} {\enquote {\bibinfo {title} {On the notion of
  balance of a signed graph.}}\ }\href {\doibase 10.1307/mmj/1028989917}
  {\bibfield  {journal} {\bibinfo  {journal} {The Michigan Mathematical
  Journal}\ }\textbf {\bibinfo {volume} {2}},\ \bibinfo {pages} {143--146}
  (\bibinfo {year} {1953}{\natexlab{b}})}\BibitemShut {NoStop}%
\bibitem [{\citenamefont {Aref}\ \emph {et~al.}(2020)\citenamefont {Aref},
  \citenamefont {Dinh}, \citenamefont {Rezapour},\ and\ \citenamefont
  {Diesner}}]{arefMultilevelStructuralEvaluation2020}%
  \BibitemOpen
  \bibfield  {author} {\bibinfo {author} {\bibfnamefont {Samin}\ \bibnamefont
  {Aref}}, \bibinfo {author} {\bibfnamefont {Ly}~\bibnamefont {Dinh}}, \bibinfo
  {author} {\bibfnamefont {Rezvaneh}\ \bibnamefont {Rezapour}}, \ and\ \bibinfo
  {author} {\bibfnamefont {Jana}\ \bibnamefont {Diesner}},\ }\bibfield  {title}
  {\enquote {\bibinfo {title} {Multilevel structural evaluation of signed
  directed social networks based on balance theory},}\ }\href {\doibase
  10.1038/s41598-020-71838-6} {\bibfield  {journal} {\bibinfo  {journal}
  {Scientific Reports}\ }\textbf {\bibinfo {volume} {10}},\ \bibinfo {pages}
  {15228} (\bibinfo {year} {2020})}\BibitemShut {NoStop}%
\bibitem [{\citenamefont {Chen}\ \emph {et~al.}(2018)\citenamefont {Chen},
  \citenamefont {Qian}, \citenamefont {Liu},\ and\ \citenamefont
  {Sun}}]{Chen_Qian_Liu_Sun_2018}%
  \BibitemOpen
  \bibfield  {author} {\bibinfo {author} {\bibfnamefont {Yiqi}\ \bibnamefont
  {Chen}}, \bibinfo {author} {\bibfnamefont {Tieyun}\ \bibnamefont {Qian}},
  \bibinfo {author} {\bibfnamefont {Huan}\ \bibnamefont {Liu}}, \ and\ \bibinfo
  {author} {\bibfnamefont {Ke}~\bibnamefont {Sun}},\ }\bibfield  {title}
  {\enquote {\bibinfo {title} {``bridge'': Enhanced signed directed network
  embedding},}\ }in\ \href {\doibase 10.1145/3269206.3271738} {\emph {\bibinfo
  {booktitle} {Proceedings of the 27th ACM International Conference on
  Information and Knowledge Management}}},\ \bibinfo {series and number} {CIKM
  '18}\ (\bibinfo  {publisher} {Association for Computing Machinery},\ \bibinfo
  {year} {2018})\ p.\ \bibinfo {pages} {773–782}\BibitemShut {NoStop}%
\bibitem [{\citenamefont {Shubin}(1994)}]{shubinDiscreteMagneticLaplacian1994}%
  \BibitemOpen
  \bibfield  {author} {\bibinfo {author} {\bibfnamefont {M.~A.}\ \bibnamefont
  {Shubin}},\ }\bibfield  {title} {\enquote {\bibinfo {title} {Discrete
  magnetic laplacian},}\ }\href {\doibase 10.1007/BF02101702} {\bibfield
  {journal} {\bibinfo  {journal} {Communications in Mathematical Physics}\
  }\textbf {\bibinfo {volume} {164}},\ \bibinfo {pages} {259--275} (\bibinfo
  {year} {1994})}\BibitemShut {NoStop}%
\bibitem [{\citenamefont {Fanuel}\ \emph {et~al.}(2017)\citenamefont {Fanuel},
  \citenamefont {Ala{\'i}z},\ and\ \citenamefont
  {Suykens}}]{fanuelMagneticEigenmapsCommunity2017}%
  \BibitemOpen
  \bibfield  {author} {\bibinfo {author} {\bibfnamefont {Micha{\"e}l}\
  \bibnamefont {Fanuel}}, \bibinfo {author} {\bibfnamefont {Carlos~M.}\
  \bibnamefont {Ala{\'i}z}}, \ and\ \bibinfo {author} {\bibfnamefont {Johan
  A.~K.}\ \bibnamefont {Suykens}},\ }\bibfield  {title} {\enquote {\bibinfo
  {title} {Magnetic eigenmaps for community detection in directed networks},}\
  }\href {\doibase 10.1103/PhysRevE.95.022302} {\bibfield  {journal} {\bibinfo
  {journal} {Physical Review E}\ }\textbf {\bibinfo {volume} {95}},\ \bibinfo
  {pages} {022302} (\bibinfo {year} {2017})}\BibitemShut {NoStop}%
\bibitem [{\citenamefont {F.~de Resende}\ and\ \citenamefont
  {da~F.~Costa}(2020)}]{chaos2020}%
  \BibitemOpen
  \bibfield  {author} {\bibinfo {author} {\bibfnamefont {Bruno~Messias}\
  \bibnamefont {F.~de Resende}}\ and\ \bibinfo {author} {\bibfnamefont
  {Luciano}\ \bibnamefont {da~F.~Costa}},\ }\bibfield  {title} {\enquote
  {\bibinfo {title} {Characterization and comparison of large directed networks
  through the spectra of the magnetic laplacian},}\ }\href {\doibase
  10.1063/5.0006891} {\bibfield  {journal} {\bibinfo  {journal} {Chaos: An
  Interdisciplinary Journal of Nonlinear Science}\ }\textbf {\bibinfo {volume}
  {30}},\ \bibinfo {pages} {073141} (\bibinfo {year} {2020})}\BibitemShut
  {NoStop}%
\bibitem [{\citenamefont {Böttcher}\ and\ \citenamefont
  {Porter}(2024)}]{complexNetWeights}%
  \BibitemOpen
  \bibfield  {author} {\bibinfo {author} {\bibfnamefont {Lucas}\ \bibnamefont
  {Böttcher}}\ and\ \bibinfo {author} {\bibfnamefont {Mason~A.}\ \bibnamefont
  {Porter}},\ }\bibfield  {title} {\enquote {\bibinfo {title} {Complex networks
  with complex weights},}\ }\href {\doibase 10.1103/PhysRevE.109.024314}
  {\bibfield  {journal} {\bibinfo  {journal} {Physical Review E}\ }\textbf
  {\bibinfo {volume} {109}},\ \bibinfo {pages} {024314} (\bibinfo {year}
  {2024})}\BibitemShut {NoStop}%
\bibitem [{\citenamefont {Peron}\ \emph {et~al.}(2020)\citenamefont {Peron},
  \citenamefont {{de Resende}}, \citenamefont {Rodrigues}, \citenamefont
  {Costa},\ and\ \citenamefont
  {{M{\'e}ndez-Berm{\'u}dez}}}]{peronSpacingRatioCharacterization2020}%
  \BibitemOpen
  \bibfield  {author} {\bibinfo {author} {\bibfnamefont {Thomas}\ \bibnamefont
  {Peron}}, \bibinfo {author} {\bibfnamefont {Bruno Messias~F.}\ \bibnamefont
  {{de Resende}}}, \bibinfo {author} {\bibfnamefont {Francisco~A.}\
  \bibnamefont {Rodrigues}}, \bibinfo {author} {\bibfnamefont {Luciano da~F.}\
  \bibnamefont {Costa}}, \ and\ \bibinfo {author} {\bibfnamefont {J.~A.}\
  \bibnamefont {{M{\'e}ndez-Berm{\'u}dez}}},\ }\bibfield  {title} {\enquote
  {\bibinfo {title} {Spacing ratio characterization of the spectra of directed
  random networks},}\ }\href {\doibase 10.1103/PhysRevE.102.062305} {\bibfield
  {journal} {\bibinfo  {journal} {Physical Review E}\ }\textbf {\bibinfo
  {volume} {102}},\ \bibinfo {pages} {062305} (\bibinfo {year}
  {2020})}\BibitemShut {NoStop}%
\bibitem [{\citenamefont {Zhang}\ \emph {et~al.}(2021)\citenamefont {Zhang},
  \citenamefont {Brugnone}, \citenamefont {Perlmutter},\ and\ \citenamefont
  {Hirn}}]{magnetNeural}%
  \BibitemOpen
  \bibfield  {author} {\bibinfo {author} {\bibfnamefont {Xitong}\ \bibnamefont
  {Zhang}}, \bibinfo {author} {\bibfnamefont {Nathan}\ \bibnamefont
  {Brugnone}}, \bibinfo {author} {\bibfnamefont {Michael}\ \bibnamefont
  {Perlmutter}}, \ and\ \bibinfo {author} {\bibfnamefont {Matthew}\
  \bibnamefont {Hirn}},\ }\bibfield  {title} {\enquote {\bibinfo {title}
  {Magnet: A magnetic neural network for directed graphs},}\ }\href
  {https://arxiv.org/abs/2102.11391v1} {\  (\bibinfo {year}
  {2021})}\BibitemShut {NoStop}%
\bibitem [{\citenamefont {de~Resende}\ \emph {et~al.}(2021)\citenamefont
  {de~Resende}, \citenamefont {Tokuda},\ and\ \citenamefont
  {Costa}}]{tab2graph}%
  \BibitemOpen
  \bibfield  {author} {\bibinfo {author} {\bibfnamefont {Bruno Messias~F.}\
  \bibnamefont {de~Resende}}, \bibinfo {author} {\bibfnamefont {Eric~K.}\
  \bibnamefont {Tokuda}}, \ and\ \bibinfo {author} {\bibfnamefont {Luciano
  da~Fontoura}\ \bibnamefont {Costa}},\ }\href
  {https://arxiv.org/abs/2110.01421v2} {\enquote {\bibinfo {title} {Unraveling
  the graph structure of tabular data through bayesian and spectral
  analysis},}\ } (\bibinfo {year} {2021})\BibitemShut {NoStop}%
\bibitem [{\citenamefont {Singer}(2011)}]{angSyncEigen}%
  \BibitemOpen
  \bibfield  {author} {\bibinfo {author} {\bibfnamefont {A.}~\bibnamefont
  {Singer}},\ }\bibfield  {title} {\enquote {\bibinfo {title} {Angular
  synchronization by eigenvectors and semidefinite programming},}\ }\href
  {\doibase https://doi.org/10.1016/j.acha.2010.02.001} {\bibfield  {journal}
  {\bibinfo  {journal} {Applied and Computational Harmonic Analysis}\ }\textbf
  {\bibinfo {volume} {30}},\ \bibinfo {pages} {20--36} (\bibinfo {year}
  {2011})}\BibitemShut {NoStop}%
\bibitem [{\citenamefont {Fanuel}\ \emph {et~al.}(2018)\citenamefont {Fanuel},
  \citenamefont {Ala{\'i}z}, \citenamefont {Fern{\'a}ndez},\ and\ \citenamefont
  {Suykens}}]{fanuelMagneticEigenmapsVisualization2018}%
  \BibitemOpen
  \bibfield  {author} {\bibinfo {author} {\bibfnamefont {Micha{\"e}l}\
  \bibnamefont {Fanuel}}, \bibinfo {author} {\bibfnamefont {Carlos~M.}\
  \bibnamefont {Ala{\'i}z}}, \bibinfo {author} {\bibfnamefont {{\'A}ngela}\
  \bibnamefont {Fern{\'a}ndez}}, \ and\ \bibinfo {author} {\bibfnamefont
  {Johan~A.K.}\ \bibnamefont {Suykens}},\ }\bibfield  {title} {\enquote
  {\bibinfo {title} {Magnetic eigenmaps for the visualization of directed
  networks},}\ }\href {\doibase 10.1016/j.acha.2017.01.004} {\bibfield
  {journal} {\bibinfo  {journal} {Applied and Computational Harmonic Analysis}\
  }\textbf {\bibinfo {volume} {44}},\ \bibinfo {pages} {189--199} (\bibinfo
  {year} {2018})}\BibitemShut {NoStop}%
\bibitem [{\citenamefont {Rozemberczki}\ \emph {et~al.}(2021)\citenamefont
  {Rozemberczki}, \citenamefont {Allen},\ and\ \citenamefont
  {Sarkar}}]{squirrelPaper}%
  \BibitemOpen
  \bibfield  {author} {\bibinfo {author} {\bibfnamefont {Benedek}\ \bibnamefont
  {Rozemberczki}}, \bibinfo {author} {\bibfnamefont {Carl}\ \bibnamefont
  {Allen}}, \ and\ \bibinfo {author} {\bibfnamefont {Rik}\ \bibnamefont
  {Sarkar}},\ }\href {\doibase 10.1093/comnet/cnab014} {\enquote {\bibinfo
  {title} {Multi-scale attributed node embedding},}\ } (\bibinfo {year}
  {2021})\BibitemShut {NoStop}%
\bibitem [{\citenamefont {Rossi}\ \emph {et~al.}(2023)\citenamefont {Rossi},
  \citenamefont {Charpentier}, \citenamefont {Di~Giovanni}, \citenamefont
  {Frasca}, \citenamefont {Gunnemann},\ and\ \citenamefont
  {Bronstein}}]{edgedir2023}%
  \BibitemOpen
  \bibfield  {author} {\bibinfo {author} {\bibfnamefont {Emanuele}\
  \bibnamefont {Rossi}}, \bibinfo {author} {\bibfnamefont {Bertrand}\
  \bibnamefont {Charpentier}}, \bibinfo {author} {\bibfnamefont {Francesco}\
  \bibnamefont {Di~Giovanni}}, \bibinfo {author} {\bibfnamefont {Fabrizio}\
  \bibnamefont {Frasca}}, \bibinfo {author} {\bibfnamefont {Stephan}\
  \bibnamefont {Gunnemann}}, \ and\ \bibinfo {author} {\bibfnamefont {Michael}\
  \bibnamefont {Bronstein}},\ }\bibfield  {title} {\enquote {\bibinfo {title}
  {Edge directionality improves learning on heterophilic graphs},}\ }\href
  {\doibase 10.48550/arXiv.2305.10498} {\  (\bibinfo {year} {2023}),\
  10.48550/arXiv.2305.10498}\BibitemShut {NoStop}%
\bibitem [{\citenamefont {Newman}(2005)}]{newmanBC}%
  \BibitemOpen
  \bibfield  {author} {\bibinfo {author} {\bibfnamefont {Mark~EJ}\ \bibnamefont
  {Newman}},\ }\bibfield  {title} {\enquote {\bibinfo {title} {A measure of
  betweenness centrality based on random walks},}\ }\href@noop {} {\bibfield
  {journal} {\bibinfo  {journal} {Social Networks}\ }\textbf {\bibinfo {volume}
  {27}},\ \bibinfo {pages} {39--54} (\bibinfo {year} {2005})}\BibitemShut
  {NoStop}%
\bibitem [{\citenamefont {Anderson}(1972)}]{Anderson_1972}%
  \BibitemOpen
  \bibfield  {author} {\bibinfo {author} {\bibfnamefont {P.~W.}\ \bibnamefont
  {Anderson}},\ }\bibfield  {title} {\enquote {\bibinfo {title} {More is
  different},}\ }\href {\doibase 10.1126/science.177.4047.393} {\bibfield
  {journal} {\bibinfo  {journal} {Science}\ }\textbf {\bibinfo {volume} {177}}
  (\bibinfo {year} {1972}),\ 10.1126/science.177.4047.393}\BibitemShut
  {NoStop}%
\bibitem [{\citenamefont {Goldenfeld}(1992)}]{Goldenfeld_1992}%
  \BibitemOpen
  \bibfield  {author} {\bibinfo {author} {\bibfnamefont {Nigel}\ \bibnamefont
  {Goldenfeld}},\ }\href@noop {} {\emph {\bibinfo {title} {Lectures on phase
  transitions and the renormalization group}}},\ Frontiers in physics\
  (\bibinfo  {publisher} {Addison-Wesley, Advanced Book Program},\ \bibinfo
  {address} {Reading, Mass},\ \bibinfo {year} {1992})\BibitemShut {NoStop}%
\bibitem [{\citenamefont {Villegas}\ \emph {et~al.}(2023)\citenamefont
  {Villegas}, \citenamefont {Gili}, \citenamefont {Caldarelli},\ and\
  \citenamefont {Gabrielli}}]{rgNature}%
  \BibitemOpen
  \bibfield  {author} {\bibinfo {author} {\bibfnamefont {Pablo}\ \bibnamefont
  {Villegas}}, \bibinfo {author} {\bibfnamefont {Tommaso}\ \bibnamefont
  {Gili}}, \bibinfo {author} {\bibfnamefont {Guido}\ \bibnamefont
  {Caldarelli}}, \ and\ \bibinfo {author} {\bibfnamefont {Andrea}\ \bibnamefont
  {Gabrielli}},\ }\bibfield  {title} {\enquote {\bibinfo {title} {Laplacian
  renormalization group for heterogeneous networks},}\ }\href {\doibase
  10.1038/s41567-022-01866-8} {\bibfield  {journal} {\bibinfo  {journal}
  {Nature Physics}\ }\textbf {\bibinfo {volume} {19}},\ \bibinfo {pages}
  {445–450} (\bibinfo {year} {2023})}\BibitemShut {NoStop}%
\bibitem [{\citenamefont {Caso}\ \emph {et~al.}(2023)\citenamefont {Caso},
  \citenamefont {Trappolini}, \citenamefont {Bacciu}, \citenamefont {Liò},\
  and\ \citenamefont {Silvestri}}]{rgGCN}%
  \BibitemOpen
  \bibfield  {author} {\bibinfo {author} {\bibfnamefont {Francesco}\
  \bibnamefont {Caso}}, \bibinfo {author} {\bibfnamefont {Giovanni}\
  \bibnamefont {Trappolini}}, \bibinfo {author} {\bibfnamefont {Andrea}\
  \bibnamefont {Bacciu}}, \bibinfo {author} {\bibfnamefont {Pietro}\
  \bibnamefont {Liò}}, \ and\ \bibinfo {author} {\bibfnamefont {Fabrizio}\
  \bibnamefont {Silvestri}},\ }\bibfield  {title} {\enquote {\bibinfo {title}
  {Renormalized graph neural networks},}\ }\href
  {http://arxiv.org/abs/2306.00707} {\  (\bibinfo {year} {2023})},\ \bibinfo
  {note} {arXiv:2306.00707 [physics]}\BibitemShut {NoStop}%
\bibitem [{\citenamefont {Nurisso}\ \emph {et~al.}(2024)\citenamefont
  {Nurisso}, \citenamefont {Morandini}, \citenamefont {Lucas}, \citenamefont
  {Vaccarino}, \citenamefont {Gili},\ and\ \citenamefont
  {Petri}}]{rgHighOrder}%
  \BibitemOpen
  \bibfield  {author} {\bibinfo {author} {\bibfnamefont {Marco}\ \bibnamefont
  {Nurisso}}, \bibinfo {author} {\bibfnamefont {Marta}\ \bibnamefont
  {Morandini}}, \bibinfo {author} {\bibfnamefont {Maxime}\ \bibnamefont
  {Lucas}}, \bibinfo {author} {\bibfnamefont {Francesco}\ \bibnamefont
  {Vaccarino}}, \bibinfo {author} {\bibfnamefont {Tommaso}\ \bibnamefont
  {Gili}}, \ and\ \bibinfo {author} {\bibfnamefont {Giovanni}\ \bibnamefont
  {Petri}},\ }\bibfield  {title} {\enquote {\bibinfo {title} {Higher-order
  laplacian renormalization},}\ }\href {\doibase 10.48550/arXiv.2401.11298} {\
  (\bibinfo {year} {2024}),\ 10.48550/arXiv.2401.11298},\ \bibinfo {note}
  {arXiv:2401.11298 [cond-mat, physics:physics]}\BibitemShut {NoStop}%
\bibitem [{\citenamefont {Loures}\ \emph {et~al.}(2023)\citenamefont {Loures},
  \citenamefont {Piovesana},\ and\ \citenamefont {Brum}}]{rgBrasil}%
  \BibitemOpen
  \bibfield  {author} {\bibinfo {author} {\bibfnamefont {Matheus de~C.}\
  \bibnamefont {Loures}}, \bibinfo {author} {\bibfnamefont {Alan~Albert}\
  \bibnamefont {Piovesana}}, \ and\ \bibinfo {author} {\bibfnamefont
  {José~Antônio}\ \bibnamefont {Brum}},\ }\bibfield  {title} {\enquote
  {\bibinfo {title} {Laplacian coarse graining in complex networks},}\ }\href
  {\doibase 10.48550/arXiv.2302.07093} {\  (\bibinfo {year} {2023}),\
  10.48550/arXiv.2302.07093}\BibitemShut {NoStop}%
\bibitem [{\citenamefont {Adamic}\ and\ \citenamefont
  {Glance}(2005)}]{polblogs2005}%
  \BibitemOpen
  \bibfield  {author} {\bibinfo {author} {\bibfnamefont {Lada~A.}\ \bibnamefont
  {Adamic}}\ and\ \bibinfo {author} {\bibfnamefont {Natalie}\ \bibnamefont
  {Glance}},\ }\bibfield  {title} {\enquote {\bibinfo {title} {The political
  blogosphere and the 2004 u.s. election: divided they blog},}\ }in\ \href
  {\doibase 10.1145/1134271.1134277} {\emph {\bibinfo {booktitle} {Proceedings
  of the 3rd international workshop on Link discovery}}}\ (\bibinfo
  {publisher} {ACM},\ \bibinfo {address} {Chicago Illinois},\ \bibinfo {year}
  {2005})\BibitemShut {NoStop}%
\bibitem [{\citenamefont {MacKay}\ \emph {et~al.}(2020)\citenamefont {MacKay},
  \citenamefont {Johnson},\ and\ \citenamefont
  {Sansom}}]{mackayHowDirectedDirected2020}%
  \BibitemOpen
  \bibfield  {author} {\bibinfo {author} {\bibfnamefont {R.~S.}\ \bibnamefont
  {MacKay}}, \bibinfo {author} {\bibfnamefont {S.}~\bibnamefont {Johnson}}, \
  and\ \bibinfo {author} {\bibfnamefont {B.}~\bibnamefont {Sansom}},\
  }\bibfield  {title} {\enquote {\bibinfo {title} {How directed is a directed
  network?}}\ }\href {\doibase 10.1098/rsos.201138} {\bibfield  {journal}
  {\bibinfo  {journal} {Royal Society Open Science}\ }\textbf {\bibinfo
  {volume} {7}},\ \bibinfo {pages} {201138} (\bibinfo {year}
  {2020})}\BibitemShut {NoStop}%
\bibitem [{\citenamefont {Moutsinas}\ and\ \citenamefont
  {Shuaib}()}]{graphHierarchy2021}%
  \BibitemOpen
  \bibfield  {author} {\bibinfo {author} {\bibfnamefont {Giannis}\ \bibnamefont
  {Moutsinas}}\ and\ \bibinfo {author} {\bibfnamefont {Choudhry}\ \bibnamefont
  {Shuaib}},\ }\bibfield  {title} {\enquote {\bibinfo {title} {Graph hierarchy:
  A novel framework to analyse hierarchical structures in complex networks},}\
  }\href@noop {} {\ ,\ \bibinfo {pages} {13}}\BibitemShut {NoStop}%
\bibitem [{\citenamefont {Gong}\ \emph {et~al.}()\citenamefont {Gong},
  \citenamefont {Higham},\ and\ \citenamefont
  {Zygalakis}}]{directedNetworkLaplacians2021}%
  \BibitemOpen
  \bibfield  {author} {\bibinfo {author} {\bibfnamefont {Xue}\ \bibnamefont
  {Gong}}, \bibinfo {author} {\bibfnamefont {Desmond~J.}\ \bibnamefont
  {Higham}}, \ and\ \bibinfo {author} {\bibfnamefont {Konstantinos}\
  \bibnamefont {Zygalakis}},\ }\bibfield  {title} {\enquote {\bibinfo {title}
  {Directed network laplacians and random graph models},}\ }\href {\doibase
  10.1098/rsos.211144} {\bibfield  {journal} {\bibinfo  {journal} {Royal
  Society Open Science}\ }\textbf {\bibinfo {volume} {8}},\ \bibinfo {pages}
  {211144}}\BibitemShut {NoStop}%
\bibitem [{\citenamefont {{Jianbo Shi}}\ and\ \citenamefont
  {{Malik}}(2000)}]{normalizedCutImgSeg2000}%
  \BibitemOpen
  \bibfield  {author} {\bibinfo {author} {\bibnamefont {{Jianbo Shi}}}\ and\
  \bibinfo {author} {\bibfnamefont {J.}~\bibnamefont {{Malik}}},\ }\bibfield
  {title} {\enquote {\bibinfo {title} {Normalized cuts and image
  segmentation},}\ }\href@noop {} {\bibfield  {journal} {\bibinfo  {journal}
  {IEEE Transactions on Pattern Analysis and Machine Intelligence}\ }\textbf
  {\bibinfo {volume} {22}},\ \bibinfo {pages} {888--905} (\bibinfo {year}
  {2000})}\BibitemShut {NoStop}%
\bibitem [{\citenamefont {Casaca}\ \emph {et~al.}(2011)\citenamefont {Casaca},
  \citenamefont {Paiva},\ and\ \citenamefont {Nonato}}]{spectralCartoonNonato}%
  \BibitemOpen
  \bibfield  {author} {\bibinfo {author} {\bibfnamefont {Wallace}\ \bibnamefont
  {Casaca}}, \bibinfo {author} {\bibfnamefont {Afonso}\ \bibnamefont {Paiva}},
  \ and\ \bibinfo {author} {\bibfnamefont {Luis~Gustavo}\ \bibnamefont
  {Nonato}},\ }\bibfield  {title} {\enquote {\bibinfo {title} {Spectral
  segmentation using cartoon-texture decomposition and inner product-based
  metric},}\ }in\ \href@noop {} {\emph {\bibinfo {booktitle} {2011 24th
  SIBGRAPI Conference on Graphics, Patterns and Images}}}\ (\bibinfo
  {organization} {IEEE},\ \bibinfo {year} {2011})\ pp.\ \bibinfo {pages}
  {266--273}\BibitemShut {NoStop}%
\end{thebibliography}%
	\cleardoublepage
	\appendix

\section{ The HodgeRank solution and the relationships with SpringRank and  Trophic Laplacian}
The HodgeRank approach has been rediscovered across the years using different
approaches. As example   \cite{debaccoPhysicalModelEfficient2018} derives a
ranking method a.k.a. \textit{SpringRank} by using a physically inspired model.
Specifically, they proposed a ranking position function,
$s: V\rightarrow \mathbb R_+$, that allows a energy being associated with the graph. Such energy function is given by the following equation

\begin{eqnarray}
H(s)  = \frac{1}{2} \sum_{u, v\in V}A(u, v)(s(u) -s(v) -1)^2.
\label{eqSprinRankHamil}
\end{eqnarray}

Therefore, due the convexity they propose to find the $s$ such as the bellow equation is 
satisfied
\begin{align}
\left(
\mathbf D^{in}+ \mathbf D^{out} - (\mathbf A+\mathbf A.T)
\right)\mathbf s^* = (\mathbf D^{out}- \mathbf D^{in})\mathbf 1.
\label{eqSpringRank}
\end{align}
With the above equation and the assumptions proposed in
\cite{debaccoPhysicalModelEfficient2018} we can find the connections with the
SprinRank method and the  Hodge-Helmholtz decomposition and consequently with
the HodgeRank method. 
First, notice that the left hand side is just the combinatorial Laplacian
associated with a unweighted graph.  In addition, 
the right hand side is divergence of the combinatorial Laplacian
applied into the flux $A-A^T$. Thus the
\eqreff{eqSpringRank} can be rewritten as
\begin{align}
\left(
\Delta_0 s^\star 
\right)(u) &= \sum\limits_{v\in V} (w(v, u)- w(u, v))\nonumber\\
&=-(\mathrm{div} F)(u).
\end{align}
The above expression is the same presented in Theorem 3 by
\cite{jiangStatisticalRankingCombinatorial2011} in the absence of weighted preferences. Consequently the solution
of the HodgeRank and SpringRank are exact the same. Henceforth, the
interpretation is also similar. The Helmholtz potential obtained by the
Hodge-decomposition allows to associate a gradient flow, $F_g$, to a directed
graph. This flow has the same meaning that obtained by the
\eqreff{eqSprinRankHamil}. That is, the flow is due a gradient of a potential
function, $\phi$. In ensence \eqreff{eqSpringRank} represents the unweighted
version of the Hodge-rank method.

Recently, \cite{mackayHowDirectedDirected2020} derived a method a.k.a .as
trophic levels defining a Trophic Laplacian, $\Lambda$.
This method has been used in applications from biology analysis to financial
systems\cite{graphHierarchy2021, mackayHowDirectedDirected2020} and further 
generative models has been proposed  \cite{directedNetworkLaplacians2021}.
However, is noticeable that the Trophic Laplacian is the same that 
combinatorial Lalplacian for $G_s$. Hence, the expression for $\mathbf \Lambda$
is exact the same that presented in the left hand side of\eqreff{eqSpringRank}. 
We can notice that the only difference between the SpringRank and Trophic methods
are that the latter looks for a $h: V\rightarrow \mathbb R_+$ such as the following
equation is respected
\begin{align}
\left(
\mathbf D^{in}+ \mathbf D^{out} - (\mathbf A+\mathbf A.T)
\right)\mathbf h^* = (\mathbf D^{in}- \mathbf D^{out})\mathbf 1.
\label{eqTrophic}
\end{align}

Which is Hodge-Helmholtz decomposition with a reverse flux, $A^{T} -A$.

\begin{align}
\left(
\Delta_0 s^\star 
\right)(u) &= \sum\limits_{v\in V} (w(v, u)- w(u, v))\nonumber\\
&=-(\mathrm{div} (-F))(u).
\end{align}

The relationships between the Laplacian operators and functions discussed in
this appendix are summarized in Table \ref{tableHodgeRankSpringRankTrophic}.

\begin{table*}[ht!]
	\caption{Relationships between the concepts in the HodgeRank, SpringRank and Trophic Laplacian methods.}
	\label{tableHodgeRankSpringRankTrophic}
	\vskip 0.15in
	\begin{center}
		\begin{small}
			\begin{sc}
				\begin{tabular}{lcccrcc}
					\toprule
					HodgeRank & SpringRank & Trophic Laplacian \\
					\midrule
					Helmholtz Potential ($\phi$) & Node Positional rank ($s$) & trophic levels ($h$) \\
					Flow divergence ($\nabla F$) & $D_{out}
					-D_{in}$ & node imbalances ($v$)\\
					Combinatorial Laplacian ($\Delta_0$)  & $D_{out}
					+D_{in} -(A + A^T)$ & Trophic Laplacian ($\Lambda$)\\
					Residual &  $\frac{1}{2} \sum_{u, v\in V}A(u, v)(s(u) -s(v) -1)^2$ 
					& 
					 $\frac{
					 	\sum_{m,n}
					 	w_{mn}(h_n -h_m -1)^2
					}
				 	{w_{mn}}$ \\
					\bottomrule
				\end{tabular}
			\end{sc}
		\end{small}
	\end{center}
	\vskip -0.1in
\end{table*}
\section{Image to Directed Graph}
\label{sec:appendixImg2DiGraph}

\subsection{Symbols}

\begin{itemize}
	\item $V:$ the set of pixels.
	\item $|V|=N_p\in \mathbb N:$ number of pixels.
	\item $C:$ the original image.
	\item $\mathcal C:$ the $C$ image  with all pre-processing steps applied. Here, we applied just the downscale in the image using a bi-cubic interpolation in order to reduce the computational costs when it is necessary.
	\item $I(u)\in[0,1]:$ the intensity pixel function.
	\item $p:$ the pixel position function, which maps a pixel into a pixel coordinate.
	\item $q\in[0, 1]:$ the charge parameter related with the magnetic formalism.
	\item $d_P:$ the metric  function in the space of pixels position. Here we just use the Euclidean distance.
	
	\item $\mathcal Nei(u)\subset V:$ the function which maps a pixel into his  neighborhood.
	
	\item $\sigma_I> 0$ (used in kernel approaches) position variance in the exponential function.
	\item $\sigma_p> 0$ (used in kernel approaches)intensity variance in the exponential function.
	\item $\Delta I_{min}\in \mathbb R_+$ (used in kernel approaches) the minimum intensity difference in order to add a directed edge.
	
	\item $\alpha \in \mathbb R_+$ (used in kernel approaches) this parameter controls how fast the hyperbolic tangent grows in the kernel approach.

	\item $\eta \in \mathbb R_+$ (used in gradient approach) this parameter controls how fast the weight decreases when the directional derivative  grows.
	\item $d_{uv}$ (used in gradient approach) the displacement vector bettwen the pixels $u$ and $v$, $d_{uv} = \frac{p(v)-p(u)}{||p(v)-p(u)||_P}$.
\end{itemize}
\subsection{Image to DiGraph}

Here we propose some methods in order to map images into directed graphs. Some of them are inspired in the methods for undirected graphs construction.
\subsubsection{Kernel approach}

A common method used in the literature to build undirected graphs associated with a given image is to use a exponential weight function\cite{normalizedCutImgSeg2000} which gives the strength of the connection between the pixels belonging to the same pre-defined neighborhood. That is, for a given $u$ and for each $v\in \mathcal Nei(u)$ we have an edge with the following weight value
\begin{align}
	w(u,v) = &e^{
		-\frac{||p(u) - p(v)||_P^2}{\sigma_s}
		-\frac{|I(u) - I(v)|}{\sigma_I}
	},
\end{align}
where $\sigma_s, \sigma_I > 0$.

In order to build an analog  for directed graphs we propose  just incorporate as a direction of the edge the signal of $I(u) - I(v)$. That is, for a given $\Delta I_{min}\ge 0$ and for each $u$ and $v$ which are neighbors, we add an (undirected or directed) edge with weight $w(u, v)$ using the following conditions
\begin{itemize}
	\item If  $I(v)-I(u) > I_{min}$
	the edge is directed and goes from $u$ to $v$.
	\item If  $I(v)-I(u) < -I_{min}$
	the edge is directed and goes from $v$ to $u$.
	\item Otherwise 
	the edge it is undirected.
\end{itemize}

The construction explained above have an simple connection with the undirected approaches. However, for reasons we will explain latter we also will study a other kernel where the weight function for directed edges are given by the following function
\begin{align}
	w_\alpha(u,v) = & \tanh (\alpha |I(u)-I(v)|),
\end{align}
where $\alpha \in \mathbb R_+$.

\subsubsection{Gradient approach}
\label{sec:img2digraphGrad}

In the work made by Nonato et. al\cite{spectralCartoonNonato} the authors proposed to use an inner product between the gradient image (which could be calculated using the Sobel filter) and the displacement vector between the pixels in order to incorporate some information about the direction of the changes between the intensities. 

In essential, the authors defined the following weight  function 
\begin{align}
	w(u, v) = \frac{1}{1+\eta g^2(u, v)},
\end{align}
where $\eta \ge 0$ and $g$ it is given by
\begin{align}
	g(u, v) = \max \left\{
	\nabla C(v). \vec d_{uv},\ \
	\nabla C(u). \vec d_{uv},\  \
	0
	\right\}
	\label{eqG}
\end{align}
As can be noticed, the above equation is symmetric, $g(u, v)  = g(v, u)$, by construction. The authors\cite{spectralCartoonNonato} argues the symmetry it is essential in order to guarantee an set of real eigenvalues with the graph. As we told before, that it not necessary condition if we study the directed graph using some deformed Laplacian operator. 

In this work, we incorporate the the direction removing one of the  restrictions of Eq.~\eqref{eqG},
\begin{align}
	h(u, v) = \max \left\{
	\nabla C(x). \vec d_{uv},\ \
	0
	\right\},
\end{align}
and keeping the weight function
\begin{align}
	w_h(u, v)= \frac{1}{1+\eta h^2(u, v)}
\end{align}
which now, it is not necessary symmetric.


\end{document}